\documentclass[aps,prd,showpacs,showkeys,preprintnumbers]{revtex4}
\usepackage{graphicx}                              
\usepackage{amsmath,amsfonts}
\graphicspath{{./pictures/}}                       

\usepackage{epsfig}
\usepackage{amssymb}
\usepackage{amsfonts}
\usepackage{float}
\usepackage{bbm}
\usepackage{psfrag}
\newcommand{\vs}{\vspace}
\newcommand{\hs}{\hspace}

\newcommand{\bdm}{\begin{displaymath}}
\newcommand{\edm}{\end{displaymath}}
\newcommand{\beq}{\begin{equation}}
\newcommand{\eeq}{\end{equation}}
\newcommand{\bea}{\begin{eqnarray}}
\newcommand{\eea}{\end{eqnarray}}
\newcommand{\bit}{\begin{itemize}}
\newcommand{\eit}{\end{itemize}}
\newcommand{\bc}{\begin{center}}
\newcommand{\ec}{\end{center}}
\newcommand{\re}{\relax{\rm I\kern-.18em R}}
\newcommand{\ID}{\mathbbm{1}}
\newcommand{\Z}{\mathbbm{Z}}
\newcommand{\N}{\mathbbm{N}}

\newcommand{\ie}{{\it i.e. }}
\newcommand{\pol}{\ell} 
\newcommand{\polInf}{{\cal P}_{\infty}} 

\newcommand{\multi}{PA}

\begin{document}
\preprint{HU-EP-06/35}

\title{Improved superposition schemes for \\ 
approximate multi-caloron configurations} 

\author{P. Gerhold, E.-M. Ilgenfritz, and M. M\"uller-Preussker}
\affiliation{Humboldt-Universit\"at zu Berlin, Institut f\"ur Physik, 
Newtonstr. 15, D-12489 Berlin, Germany}

\date{March 15, 2007}

\begin{abstract}

Two improved superposition schemes for the construction of approximate 
multi-caloron-anticaloron configurations, using exact single 
(anti)caloron gauge fields as underlying building 
blocks, are introduced in this paper. The first improvement deals with 
possible monopole-Dirac string interactions between different calorons 
with non-trivial holonomy. The second one, based on the ADHM 
formalism, improves the (anti-)selfduality in the case of small caloron 
separations. It conforms with Shuryak's well-known ratio-ansatz when applied 
to instantons. Both superposition techniques provide a higher degree of 
(anti-)selfduality than the widely used sum-ansatz, which simply adds the 
(anti)caloron vector potentials in an appropriate gauge. Furthermore, 
the improved configurations (when discretized onto a lattice) are 
characterized by a higher stability when they are exposed to 
lattice cooling techniques.
\end{abstract}



\maketitle

\section{Introduction}
\label{sec:Introduction}
Lately, calorons with non-trivial asymptotic 
holonomy~\cite{Kraan:1998kp,Kraan:1998pm,Lee:1998bb} 
(besides the longer known ones with trivial asymptotic 
holonomy, HS calorons~\cite{Harrington:1978ve}) have attracted some 
interest, both as 
solutions themselves~\cite{Bruckmann:2002vy,Bruckmann:2004nu} 
and as underlying building blocks of some semiclassical-like 
modeling of QCD in the finite temperature case~\cite{Gerhold:2006sk}
by a caloron-anticaloron gas model. 
The motivation of such simulations was to check whether these solutions,
rather than instantons, are able to produce confinement.
In some sense these solutions are better adapted to varying physical
situations including the confinement phase where the average holonomy is
non-trivial. One interesting feature of these solutions is that they can 
occur decomposed into constituents with fractional topological charge. 
As we will see, this fact causes part of the problems that require to 
reconsider the superposition of calorons.

The semiclassical picture of QCD dates back to Callan, Dashen and 
Gross (CDG)~\cite{Callan:1977gz}. 
The authors have worked out the instanton effects on the heavy quark 
potential~\cite{Callan:1978ye} and proposed a confinement 
mechanism for quarks being enclosed in bubbles (``bags'') of the wrong, 
perturbative vacuum~\cite{Callan:1978gn,Callan:1978bm}. Finally, this mechanism 
could not be verified by later research. It has become also an accepted fact, 
that the instanton gas or liquid is unable to generate a confining potential 
between static color charges. The original proposal of CDG, however, to explain 
confinement by merons~\cite{Callan:1977qs} is fascinating people working on 
confinement until today~\cite{Negele:2004hs}. Merons are sort of singular 
half-instantons eventually liberated by a phase transition from weak to 
strong coupling. Each of them is carrying half-integer topological charge. 
Since then, the concept of fractional topological charges (``instanton quarks''
of charge $Q=1/N_{\rm color}$) is vivid in the literature 
(see Refs.~\cite{Jaimungal:1999sv,Jaimungal:1999wn} and 
Ref.~\cite{Zhitnitsky:2006sr} and references therein). Some indications
for the existence of smooth $Q=1/2$ lumps in $SU(2)$ lattice 
configurations and/or their relation to confinement have been published in the 
90-s~\cite{GarciaPerez:1993ab,GarciaPerez:1993jw,Gonzalez-Arroyo:1995zy,Gonzalez-Arroyo:1995ex,Gonzalez-Arroyo:1996jp,Edwards:1998dj} 
and more recently~\cite{Bruckmann:2004ib}. The corresponding solutions for zero 
temperature are still elusive.  

The application of instanton physics to finite temperature QCD was mainly 
influenced by the seminal paper~\cite{Gross:1980br}. Therein, for some reasons, 
the potential significance of Euclidean solutions with non-trivial holonomy has 
been overlooked. Therefore the instanton model (for a general review see 
Ref.~\cite{Schafer:1996wv}) was applied to $T \ne 0$ exclusively based on HS 
calorons~\cite{Schafer:1995pz}. Diakonov has recently pointed 
out~\cite{Diakonov:2002fq} that these arguments are not compelling for 
dense multi-caloron-anticaloron systems. His paper has stimulated more attention 
to these new caloron solutions. Ideas about caloron constituents as relevant 
degrees of freedom for the confined phase from the point of view of 
electric-magnetic duality have been developed in Ref.~\cite{Bruckmann:2005bc}.

The encouragement for the simulation of gases (or liquids) of calorons and 
anticalorons, apart from the motivation described above, came from lattice 
studies. Soon after the discovery of the caloron solutions 
Garcia Perez {\it et al.}~\cite{GarciaPerez:1999hs,GarciaPerez:1999bc}
have considered such objects on a finite lattice, partly with twisted boundary 
conditions. Later on, in 
Refs.~\cite{Ilgenfritz:2002qs,Gattringer:2002tg,Ilgenfritz:2004zz,Ilgenfritz:2006ju} 
the possibilities to detect calorons in generic Monte Carlo configurations
have been developed further, in a similar spirit as earlier instanton (caloron) 
searches at zero (non-zero) 
temperature~\cite{Chu:1994vi,Chu:1994xz,GarciaPerez:1998ru}.
In view of this development it was natural to develop the tools needed 
to model multicaloron systems. The necessity of simulating such systems reflects 
the obvious fact that a strictly semiclassical calculation, which has been done 
for a single caloron~\cite{Diakonov:2004jn}, is not feasible in a strongly 
coupled system like QCD near to the deconfining temperature.

The mathematical-physical interest in the solutions themselves has led to 
remarkable success~\cite{Bruckmann:2002vy,Bruckmann:2004nu} in recent years 
concerning the construction and description of selfdual gauge fields with higher 
topological charge $Q$. The parametrizations for $|Q|\neq 1$ are nevertheless
restricted to cases of special symmetry. A general parametrization is still lacking.
Apart from this, the modeling of gauge field trajectories of QCD in a 
semiclassical-like fashion in terms of calorons is not using (like earlier 
instanton liquid models~\cite{Shuryak:1992ke,Fukushima:1996yn,Fukushima:2000ix})
exact solutions of arbitrary topological charge but approximate, only piecewise 
perfect solutions. The approximate solutions should be superpositions of elementary 
building blocks with $Q=\pm 1$. Of course, there is necessity and room to 
perfectionate the superposition scheme such that the action and the topological 
charge of purely selfdual superpositions would be additive to a high precision. 
This is not the case for the simple sum-ansatz in terms of the vector potentials 
$A^{(i)}_{\mu}(x)$ of single calorons and/or anticalorons, except for an 
infinitely dilute gas. For the case of calorons, this improvement is addressed 
here for the first time.

The first problem of any superposition of calorons with definite holonomy 
leads us beyond the simple sum-ansatz in terms of the vector potentials 
$A^{(i)}_{\mu}(x)$ (with $i$ labelling the individual (anti)calorons). 
There is the so-called algebraic gauge where the additive superposition 
is indeed possible. The gauge transformation to the periodic gauge, which 
is the more appropriate one to build the ensemble of model configurations, 
requires all of them to have the same asymptotic holonomy~\cite{Bruckmann:2002vy}. 
Once this is adopted, a reasonable model of QCD can be 
constructed~\cite{Gerhold:2006sk}.

We introduce in the present paper two new techniques to facilitate the further 
improvement of such superpositions.  One of them is specific for calorons with 
non-trivial holonomy, which may exist in a state ``dissociated'' into monopole 
constituents. These monopoles are connected by non-Abelian strings of strongly 
peaked, but regular vector potential with low action density. They turn 
into usual Dirac strings, being Abelian and singular, when the monopole separation 
tends to infinity. For simplicity we will denote the regular, non-Abelian string
as  ``Dirac string''. The presence  of such ``Dirac strings'' creates 
problems whenever one attempts to add another caloron~\cite{Bruckmann:2002vy}. 
Since the number of constituent monopoles in a caloron is equal to the number of 
colors, $N_{\rm color}$, the case of $SU(2)$ gluodynamics with two constituents 
and one Dirac string per caloron is a convenient testing ground for this type 
of problems.

The other improvement problem is connected with caloron interactions arising 
when constituents (lumps of action) happen to come too close to each other 
when the collective coordinates are sampled. For these cases, the ADHM 
formalism~\cite{Atiyah:1978ri} offers the necessary intuition for the 
improvement of the superposition.

We should remark that the quest of improvement is restricted only to 
semiclassical-like simulations. Sampling configurations with merons and 
instantons in the regular gauge~\cite{Negele:2004hs} or with more general 
pseudoparticle fields\cite{Wagner:2006qn} do not need to approximate these
by solutions of the Euclidean equations of motion. Superpositions in this case
have to be subjected to importance sampling which induces strong correlations. 
It is these long range correlations which generates confinement in this class 
of models. Our approach is complementary in the sense that our superpositions 
are much closer to real solutions and can be randomly sampled in moduli 
space~\cite{Gerhold:2006sk}. The long range correlation required for 
confinement in this case is due to the non-trivial holonomy tail of the 
gauge potentials.    

This paper is organized as follows. 
In Section~\ref{sec:Solutions} we briefly recall the properties of single 
(anti)calorons. In Section~\ref{sec:NaiveSuperpositions} we explain 
under which circumstances, in the algebraic gauge, an additive superposition 
is possible. There we also introduce technical details like the use 
of the lattice discretization and some observables characterizing the quality 
of the constructed superposition. We explain $\epsilon$-improved cooling 
for later use in this paper. In Section~\ref{sec:DiracString} we propose a 
solution for the obstruction problem presented by Dirac strings and demonstrate 
the effect for a particular two-caloron superposition. 
In Section~\ref{sec:PseudoADHM} we describe the ADHM-inspired improvement 
for closely overlapping caloron constituents. 
The success is demonstrated for a particular four-caloron superposition. 
In Section~\ref{sec:Conclusion} we summarize and draw conclusions for further work.

\section{Classical solutions of $SU(2)$ Yang-Mills theory}
\label{sec:Solutions}

The instanton, discovered in 1975~\cite{Belavin:1975fg}, is a classical 
solution of the Euclidean Yang-Mills equation of motion in flat $\re^4$, 
representing zero temperature,  with localized action density and
one unit of topological charge. For any number of colors, the instanton is 
basically an $SU(2)$ object that is parametrized by 
$~4 \cdot N_{\rm color}~$ collective coordinates. For 
${N_{\rm color} = 2}$ these are its four-dimensional position in space-time,
a size parameter and three parameters describing a global rotation in $SU(2)$ 
group space. For higher $N_{\rm color}$ the additional parameters describe 
the embedding of $SU(2)$ into $SU(N_{\rm color})$. Classical solutions with 
higher topological charge can in principle be constructed by means of the ADHM 
formalism~\cite{Atiyah:1978ri}, although they are not analytically available 
in general. 

The analog of the instanton in the case of non-zero temperature 
$~T \equiv 1/\beta \neq 0~$
was discovered soon after the $T=0$ instanton by Harrington and Shepard in 
1978~\cite{Harrington:1978ve} and was accepted as an appropriate semiclassical 
background at finite temperature. It will be referred to as `HS caloron' 
throughout the present paper. 

Twenty years later, Kraan and van Baal~\cite{Kraan:1998kp,Kraan:1998pm} and 
Lee and Lu~\cite{Lee:1998bb} extended the parameter space of the HS caloron 
by an additional parameter, the asymptotic holonomy 
${\cal P}_{\infty}$, which is defined as
\beq 
{\cal P}_{\infty} = e^{2\pi i \vec \omega \vec \tau} = 
\lim\limits_{|\vec x| \rightarrow \infty} P(\vec x) \; ,
\label{eq:holonomy}
\eeq
where $\vec\tau$ denotes the vector of Pauli matrices and 
\beq 
P(\vec x) = \hat P\, \exp\left( i \int\limits_0^\beta A_4(\vec x,t) dt\right) 
\in SU(2) \; ,  
\label{eq:Polyakov}
\eeq 
where path ordering is implied by the $\hat P$ symbol. According to this
definition the Polyakov loop $~\pol(\vec x) \equiv \frac{1}{2} Tr\,P(\vec x)~$ 
at spatial infinity is directly determined by the parameter $\vec \omega$.
The generalized solution, which we will call KvBLL caloron, is reduced to the 
HS caloron for trivial asymptotic holonomy, taking values in the center of the 
gauge group, ${\cal P}_{\infty} \in Z(2)$. In the case of general 
$N_{\rm color}$, this solution can also be constructed~\cite{Kraan:1998sn}.
Generically, 
it is not a simple embedding of an $SU(2)$ solution into the bigger group. 
The present paper, however, is restricted to the case of $N_{\rm color}=2$.

The KvBLL caloron with arbitrary asymptotic holonomy is a selfdual gauge field, 
{\it i.e.} ${F_{\mu\nu}=\tilde{F}_{\mu\nu}}$ with 
${\tilde{F}_{\mu\nu}=\frac{1}{2} \epsilon_{\mu\nu\alpha\beta} F_{\alpha\beta}}$,
solving the Euclidean equation of motion at finite temperature. It was first 
constructed in the so-called algebraic gauge by means of the ADHM 
formalism~\cite{Atiyah:1978ri} by Kraan and van 
Baal~\cite{Kraan:1998kp,Kraan:1998pm} and by Lee and Lu~\cite{Lee:1998bb} 
as a selfdual gauge field in flat $\re^4$ with the periodicity condition 
$A^{alg}_\mu(x+\beta \hat e_t)= {\cal P}_{\infty} A^{alg}_\mu(x) 
{\cal P}_{\infty}^{\dagger}$, 
where $\hat e_t$ denotes the unit vector along the time direction. 
The resulting vector potential $A_\mu^{alg}$ defined on $\re^4$ can then 
be transformed into a selfdual and time-periodic gauge field $A_\mu^{per}$, 
defined on $\re^3 \times S^1$, by a gauge transformation $\Omega(x)$
non-periodic in time, according to
\beq
A_\mu^{per}(\vec x, t\, \mbox{mod}\, \beta) = 
        \Omega(x) A^{alg}_\mu(\vec x,t)\Omega^\dagger(x) 
   - i\,\Omega(x)\partial_\mu\Omega^\dagger(x) \; ,
\label{eq:eq30F}
\eeq
with $x=(\vec x, t)$ and
\beq
\Omega(x) = e^{- 2\pi i\vec\tau \vec\omega t/\beta} \; . 
\label{eq:Omega}
\eeq

This is the KvBLL caloron in the periodic gauge with arbitrary asymptotic 
holonomy ${{\cal P}_{\infty} = \exp(2\pi i \vec \omega\vec\tau) \in SU(2)}$
and topological charge $~Q = 1$.
One finds that the KvBLL caloron is also described by 8 collective coordinates 
in addition to its holonomy parameter $\vec \omega$. However, the physical 
interpretation of these collective coordinates is different from that of the 
instanton parameters, as will be discussed in the following.

One can imagine the KvBLL caloron in the algebraic gauge as an infinite chain 
of equally separated, identical instantons in flat $\re^4$, aligned along the 
time direction with each subsequent instanton rotated by the 
holonomy ${\cal P}_{\infty}$ relative to the preceding one - 
as long as the size $\rho$ of the instantons is small compared to the inverse 
temperature $\beta$. In that case it is appealing to parametrize the caloron in 
terms of the collective coordinates of the instanton.
For $\rho\ll\beta$ the corresponding caloron consists
of one lump of action, which is approximately $O(4)$-rotationally symmetric in 
space-time, and its radius can be described by the parameter $\rho$. 

\begin{figure*}[htbp]
\centering
\mbox{
\begin{minipage}{0.5\linewidth}
\includegraphics[angle=0,width=1.0\textwidth]{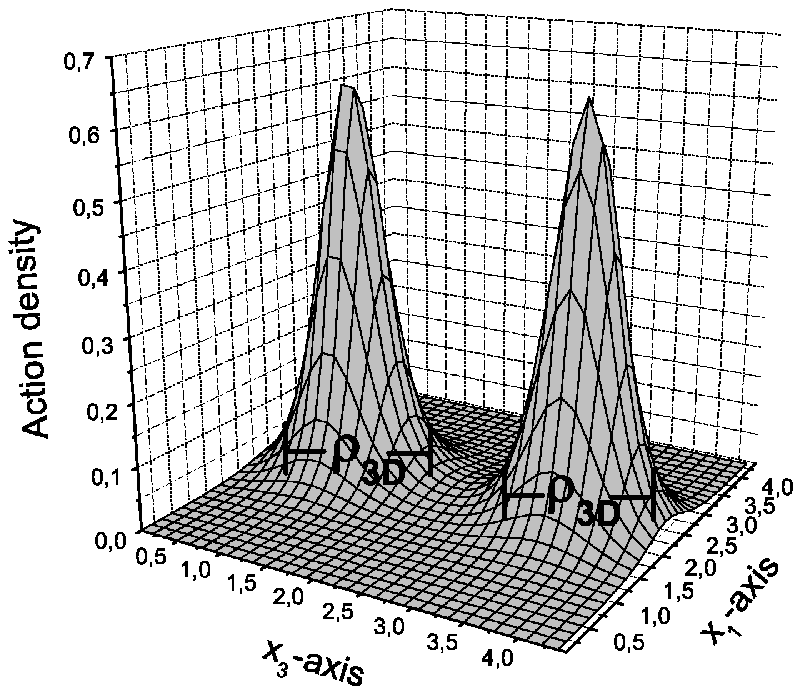}
\end{minipage}
\begin{minipage}{0.50\linewidth}
\includegraphics[angle=0,width=1.0\textwidth]{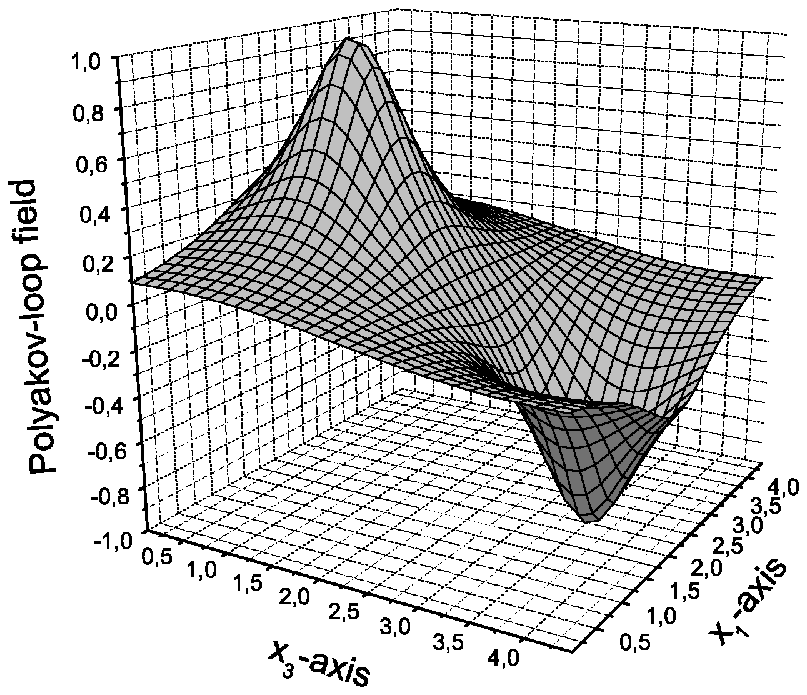}
\end{minipage}
}
\caption{Action density~(left) and Polyakov loop field $\pol(\vec x)$ (right)
of a KvBLL caloron with maximally non-trivial holonomy ${|\vec\omega|=0.25}$, 
${\rho=1.0}$ and ${\beta = 1}$, where the action density is given in instanton 
units $S_I$.
For sufficiently large $\rho/\beta$ the two constituents of the KvBLL caloron
become well separated. Here, $\rho_{3D}$ denotes the radius of the almost 
$O(3)$-symmetric monopole action lumps in 3 dimensions.}
\label{fig:30D}
\end{figure*}

In the opposite case, ${\rho \sim \beta}$ or ${\rho \gg \beta}$, the caloron is dissociated into 
two constituents, as shown in the action density plot in Figure~\ref{fig:30D}. 
For ${\rho \rightarrow \infty}$ 
the action density of these constituents becomes static in time,
breaking down the prior $O(4)$ symmetry to an 
$O(3)$-rotational symmetry in 3-dimensional position space, 
and the corresponding 3-dimensional radii $\rho_{3D}^{(1,2)}$ of the two 
constituents converge to constants approximately given by 
${\rho_{3D}^{(1)}\approx \beta/(4\omega)}$ and 
${\rho_{3D}^{(2)}\approx\beta/(4\bar\omega)}$, respectively, where 
${\omega=|\vec \omega|\in [0;1/2]}$ and ${\bar\omega = 1/2-\omega}$. The 
constituents can be identified with BPS monopoles~\cite{Prasad:1975kr} each 
carrying the fraction $2\omega$ or $2\bar\omega$, respectively, of the total 
action and topological charge. The distance between the monopole positions 
$\vec z_1$, $\vec z_2$ is given by 
$ d =|\vec z_1 - \vec z_2| = \pi\rho^2/\beta $. 
Due to the emergence of these two monopoles at larger $\rho$ it seems more 
natural to use their positions $\vec z_1$ and $\vec z_2$ for the 
parametrization of the caloron instead of the instanton parameters involved 
in the construction scheme. The corresponding vector potentials in the 
algebraic and the periodic gauge are analytically given 
in~\cite{Kraan:1998kp,Kraan:1998pm}.

\noindent
Three further remarks shall be given here.

{\em (i)} In the case of non-trivial asymptotic holonomy ${\cal P}_{\infty}$ 
the caloron gauge field does not vanish at spatial infinity. 
In the periodic gauge the $A_4$ component of the vector potential converges 
to a constant at spatial infinity whereas the remaining three components 
approach zero,
\beq
\lim\limits_{|\vec x|\rightarrow \infty}A^{per}_4(x) = 2\pi \vec\omega 
\vec\tau/\beta, \mbox{\hs{1mm}} \lim\limits_{|\vec x|\rightarrow \infty} 
A^{per}_{1,2,3}(x) = 0 \; .
\label{eq:eq30G}
\eeq
The non-vanishing vector potential will become an issue when one tries to 
superpose KvBLL calorons.

{\em (ii)} In the case ${\omega \rightarrow 0}$ or ${\bar\omega \rightarrow 0}$ 
the asymptotic holonomy becomes trivial, one constituent vanishes and the 
remaining one becomes a HS caloron. 

{\em (iii)} Finally, the anticaloron, which is the anti-selfdual analogue 
of the caloron, carrying negative topological charge, can be obtained from 
the caloron gauge field by 
\bea
A^{anti}_i (\vec x, t) &=& -A_i(-\vec x,t) \; , \quad i=1,2,3 \; , \nonumber\\
A^{anti}_4 (\vec x, t) &=& A_4(-\vec x,t) \; .
\label{eq:dualtoantidual}
\eea

\section{Superposing calorons: A first approach and its problems}
\label{sec:NaiveSuperpositions}

In general, the superposition of $N$ selfdual gauge fields 
$A_\mu^{(i)}(x)$, $i=1,...,N$, does {\it not} provide a selfdual gauge field 
configuration due to the non-linearity of the field strength $F_{\mu\nu}(x)$ 
in terms of the vector potential. However, since the non-linearity arises 
from the commutator term $[A_\mu(x), A_\nu(x)]$ appearing in the expression 
for the field strength, such a superposition has the potential of approximately
describing a selfdual gauge field, provided that at every space-time position 
$x$ the superposition is dominated by the contribution $A_\mu^{(i)}(x)$ of 
one single caloron, or that all vector potentials are almost Abelian. 

This well-known observation has been exploited in numerous semiclassical 
model calculations based on instantons in the singular gauge \cite{Callan:1977gz}
relying on the strong localization of the instanton gauge field in space-time. 
For instantons the simplest possible superposition scheme, which plainly 
consists of adding all involved vector potentials according to the 
{\it simple sum-ansatz}
\beq
A^{sum}_\mu(x) = \sum\limits_i A_\mu^{(i)}(x), \quad i=1,...,N,
\label{eq:xxx1}
\eeq
can successfully be applied as long as the instanton cores do not come too 
close to each other. 
The meaning of  the sloppy formulation `not too close' was clarified 
in~\cite{Ilgenfritz:1980vj}, establishing the quantitative criterion 
${\rho_1^2\rho_2^2/|\xi_1-\xi_2|^4 < 1/200}$ for the applicability 
of the simple sum-ansatz for the superposition of two instantons with 
sizes $\rho_1$, $\rho_2$ and positions $\xi_1$, $\xi_2$.

When dealing with calorons with non-trivial holonomy, however, this criterion 
no  longer holds, since the KvBLL caloron gauge field then does not vanish 
even far away from the monopole locations according to Eq.~(\ref{eq:eq30G}).

Applying naively the simple sum-ansatz~(\ref{eq:xxx1}) would induce
an interaction between calorons even at infinite separations.
This is the first caloron-specific problem appearing in superpositions.
It can easily be overcome by superposing the caloron vector potentials in 
the algebraic gauge, in which the potential $A_4^{alg}$ vanishes at spatial 
infinity such that the sum-ansatz does no longer yield interactions between 
infinitely far separated calorons~\cite{Bruckmann:2002vy}. On the other hand, 
the resulting gauge field is then no longer periodic. Periodicity has to be 
restored by applying a specific gauge transformation to the sum of the vector 
potentials. This leads to the modified ansatz
\beq
A^{per}_\mu(x) = e^{-2\pi i t\vec{\omega}\vec{\tau}} \cdot 
\sum_i A^{(i),alg}_\mu(x) \cdot e^{+2\pi i t\vec{\omega}\vec{\tau}} 
+ 2\pi \vec{\omega}\vec{\tau}\cdot \delta_{\mu,4} \; ,
\label{eq:xxx3}
\eeq
which we will refer to as {\it sum-ansatz} in the following, where $\beta=1$
was chosen for convenience. However, this 
superposition scheme is only valid under the condition that only calorons 
and anticalorons with identical holonomy be superposed because the gauge 
transformation necessary to bring the caloron vector potential from the 
algebraic to the periodic gauge depends on the holonomy according to 
Eqs.~(\ref{eq:eq30F},~\ref{eq:Omega}).
It would not be possible to restore the periodicity of a configuration of 
calorons first constructed in the algebraic gauge with a global gauge 
transformation, if the configuration were a superposition of calorons with 
different holonomies~\cite{Bruckmann:2002vy}.

An additional caloron-specific issue concerning the superposition of 
well-dissociated calorons with non-trivial holonomy, which cannot
be dealt with by means of the sum-ansatz~(\ref{eq:xxx3}), is due to the 
existence of the ``Dirac string'', which in our context is a string of 
strongly peaking, but regular vector potential connecting the two monopoles 
within a caloron. Despite its strength in $A_{\mu}$, the Dirac 
string is almost free of action thanks to a fine-tuning of its gauge field. 
However, there can potentially be strong interactions between this Dirac string 
and other calorons or other monopoles, respectively, spoiling the selfduality 
of the superposed configuration. A more appropriate superposition scheme 
taking care of the caloron-Dirac string interactions will be presented 
in Section~\ref{sec:DiracString}. Furthermore, similar to the case of 
instantons, the sum-ansatz also fails, if two calorons or monopoles come too 
close to each other. An improved superposition scheme for the latter case, 
leading to smaller deviations from selfduality, is introduced in 
Section~\ref{sec:PseudoADHM}.

So far, our discussion was referring completely to the continuum. 
However, for the discussion of the improved superposition schemes to be 
presented we shall find it quite instructive to apply certain lattice 
techniques to the gauge field configurations we have generated. 
For this purpose, we introduce a grid (called ``lattice'' in the following) 
and represent the gauge field $A_\mu(x)$ of the caloron ensemble by a set 
of link variables according to 
\beq
U_{x,\mu} = \hat P\,\exp\left( i\int\limits_x^{x+a\hat\mu} 
A_\mu(y) dy_\mu\right) \; ,
\label{eq:eq2Y}
\eeq
where the path ordering symbol $\hat P$ will approximately be accounted for by 
dividing each path of integration into $n$ sub-intervals with their lengths 
${a/n \ll \bar\rho}$ being small compared to the average 
caloron size, such that the vector potential inside these sub-intervals can be 
considered to be constant. Furthermore, the convergence of the link variables 
is ensured by adjusting the number of sub-intervals dynamically, corresponding 
to the variability of the gauge field, and checked by comparing the resulting 
parallel transporter for different values of $n$. 

This introduction of a lattice makes the constructed gauge field configurations 
accessible to cooling procedures~\cite{Ilgenfritz:1985dz,Teper:1985ek}. 
From the cooling history we can then 
assess the stability of the configuration under the cooling process, which 
will give information on the ``distance'' to the ``nearest'' classical 
solution in the space of all field configurations, and thus helps to evaluate 
the quality of the applied superposition scheme. Here, we will use the 
so-called $\epsilon$-cooling~\cite{GarciaPerez:1993ki,Bruckmann:2004ib} with 
$\epsilon=0$, based on an improved lattice action 
$S_{\rm impr}[A] \equiv S_{\epsilon=0}[A]$, the lattice artefacts of 
which only start in $O(a^4)$.

To estimate the importance of a constructed configuration according to its 
weight contributing to the partition function $Z$ it is interesting to 
consider its total action. We define the global quantity $\gamma$ as the 
ratio of the total lattice action $S[A]$ (in instanton units $S_I=8\pi^2/g^2$)
and the topological charge $Q[A]$ according to
\beq
\gamma = \frac{S[A]}{|Q[A]|},
\label{eq:DefOfGamma}
\eeq
where we also use an improved definition of the lattice topological charge 
$Q[A]$~\cite{Bilson-Thompson:2002jk}.

Finally, as a measure of the local violation of (anti-)selfduality we 
introduce the quantity 
\beq
\delta(x) = \sum\limits_{\mu\nu} 
 \left| F_{\mu\nu}(x) - \kappa ~\tilde{F}_{\mu\nu}(x) \right|^2 \ge 0
 \qquad \mbox{with} \quad \kappa = \mathrm{sign}(Q) =\pm 1.
\label{eq:eq4N}
\eeq
In what follows
$\delta(x)$ will directly be determined from the continuum gauge fields. 
All other results presented for the action, topological charge, and their 
densities will be measured by the lattice techniques introduced above. 
Both the quantities $\gamma$ and $\delta(x)$ reflect the deviation
from (anti-)selfduality for $\kappa=+1~(-1)$.
For purely (anti-)selfdual configurations 
$A_\mu(x)$ the local measure $\delta(x)$ is identical to zero, while the 
global measure $\gamma$ becomes unity. Of course, these definitions 
make only sense for
superpositions built exclusively out of {\it either} calorons 
{\it or} anticalorons in an attempt to construct higher charge calorons 
or anticalorons.
 
\section{Reducing the caloron-Dirac string interaction}
\label{sec:DiracString}
In this Section we deal with the interaction between the Dirac string of a 
dissociated caloron and any further (anti)calorons. Due to its 
strength the Dirac string has the potential of creating a tremendous amount of
unwanted additional action and of completely spoiling the (local) selfduality 
of a multi-caloron system created by the sum-ansatz. Certainly, this problem 
becomes most prominent if a further monopole or caloron, respectively, is placed 
between the two monopoles of a dissociated caloron, \ie on top of the Dirac 
string (see also \cite{Bruckmann:2002vy}).

We will see that this problem can be cured, at least if the constituents causing
the Dirac string are sufficiently far separated from each other, 
by applying a certain rotation in color space (not a gauge transformation) on 
the caloron located on top of the Dirac string. This result will be obtained 
from the observation that the Dirac string of a 
caloron with non-trivial holonomy is almost pure gauge for 
well-dissociated calorons, \ie the vector potential of the Dirac string 
between the monopoles can almost be gauged to zero. However, the erased 
components of the vector potential will reappear outside of the caloron. 
Finally, the improved superposition scheme will be compared with the 
sum-ansatz.

\subsection{Structure of the Dirac string vector potential}
\label{subsec:DS_structure}

\begin{figure}[htbp]
\centering
\mbox{
\begin{minipage}{0.33\linewidth}
\includegraphics[angle=0,width=1.0\textwidth]{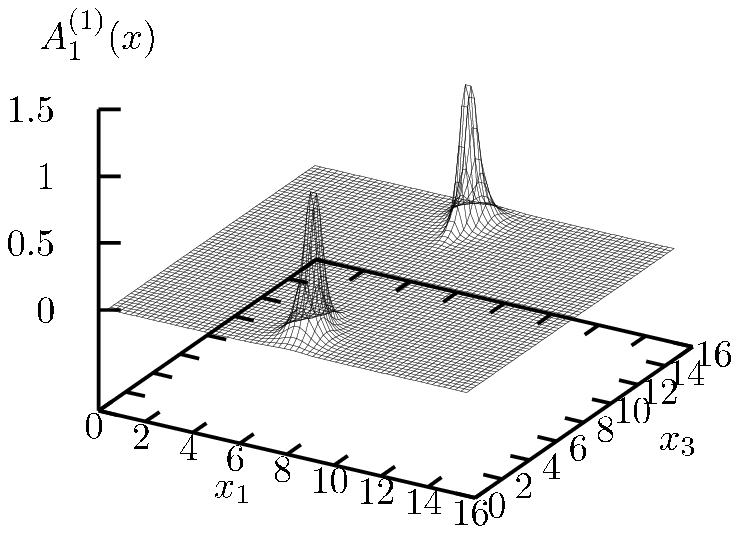}
\includegraphics[angle=0,width=1.0\textwidth]{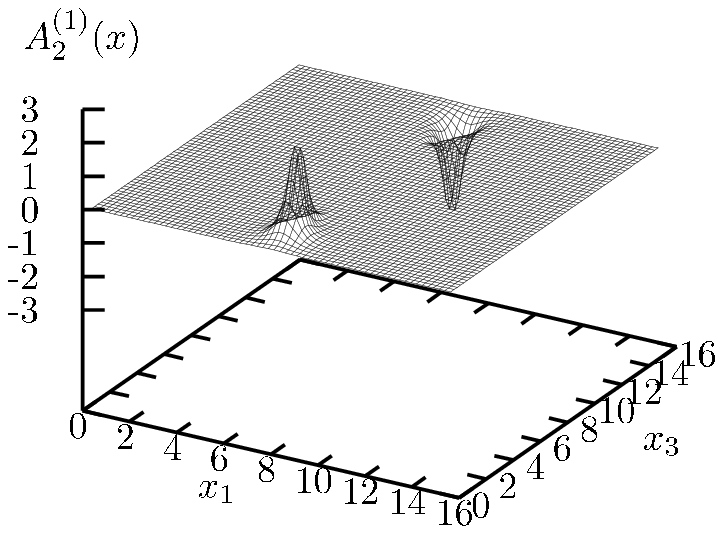}
\includegraphics[angle=0,width=1.0\textwidth]{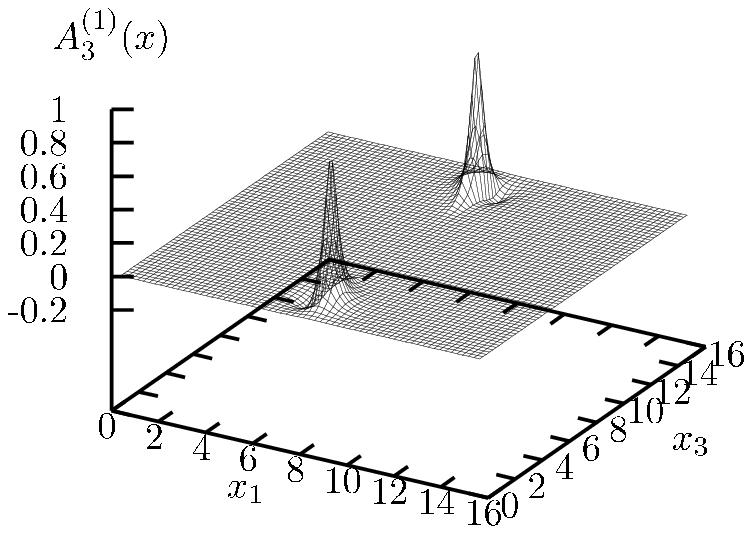}
\includegraphics[angle=0,width=1.0\textwidth]{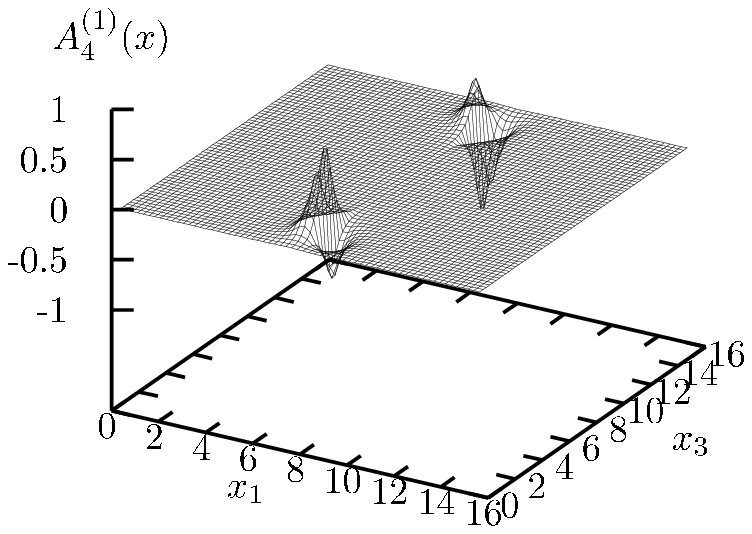}
\end{minipage}
\begin{minipage}{0.33\linewidth}
\includegraphics[angle=0,width=1.0\textwidth]{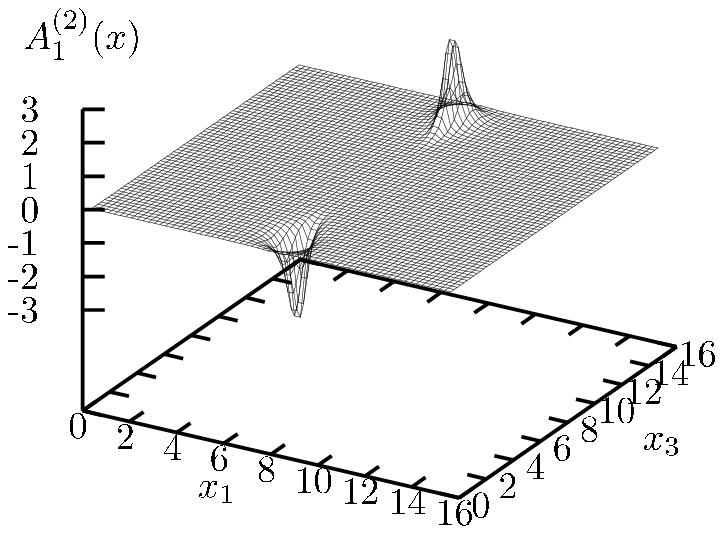}
\includegraphics[angle=0,width=1.0\textwidth]{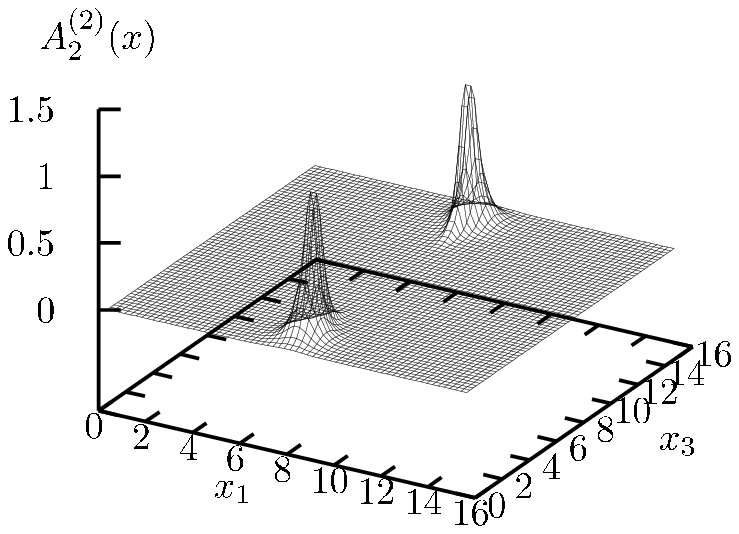}
\includegraphics[angle=0,width=1.0\textwidth]{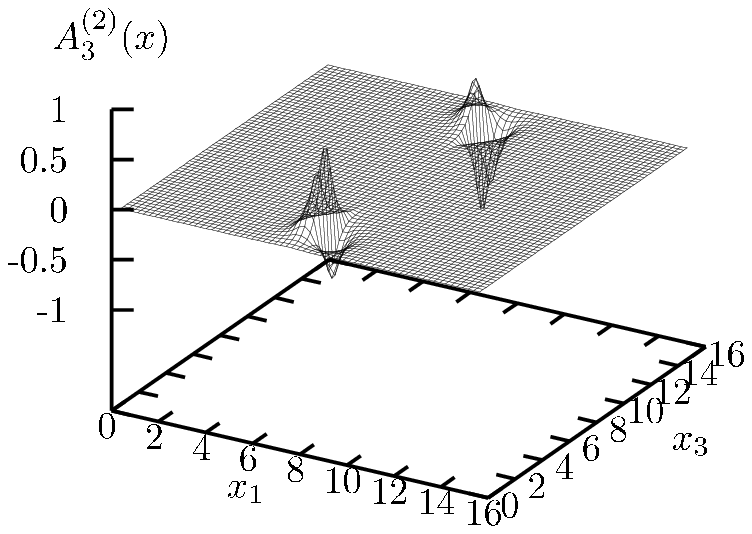}
\includegraphics[angle=0,width=1.0\textwidth]{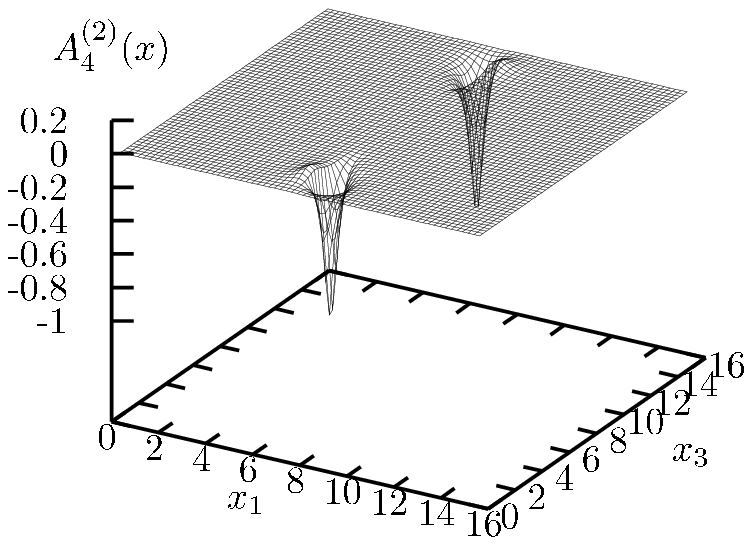}
\end{minipage}
\begin{minipage}{0.33\linewidth}
\includegraphics[angle=0,width=1.0\textwidth]{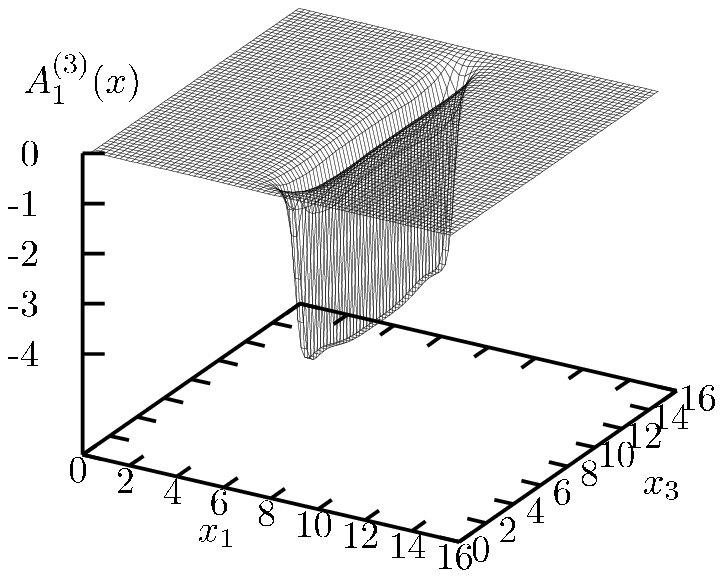}
\includegraphics[angle=0,width=1.0\textwidth]{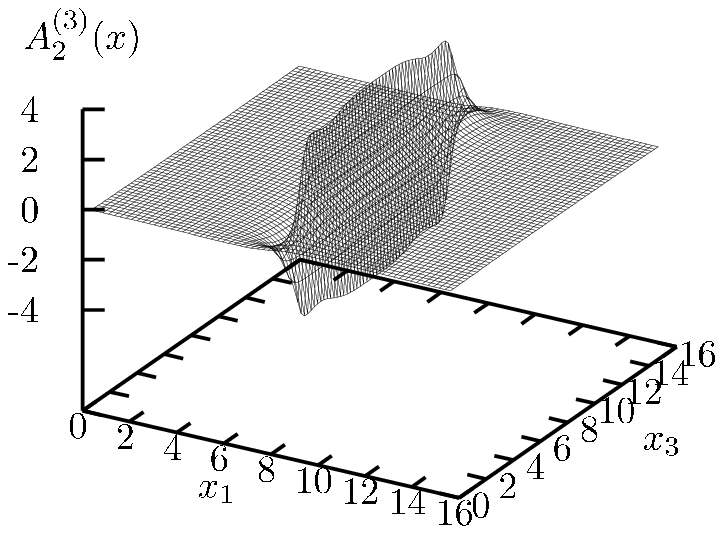}
\includegraphics[angle=0,width=1.0\textwidth]{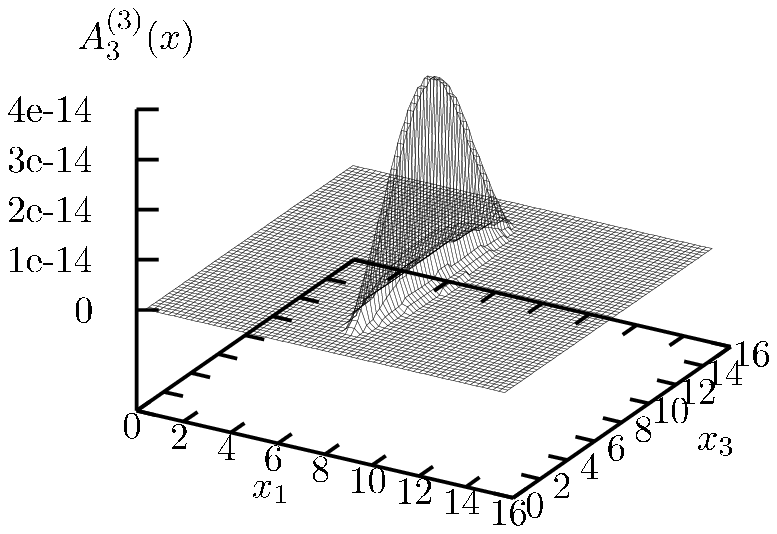}
\includegraphics[angle=0,width=1.0\textwidth]{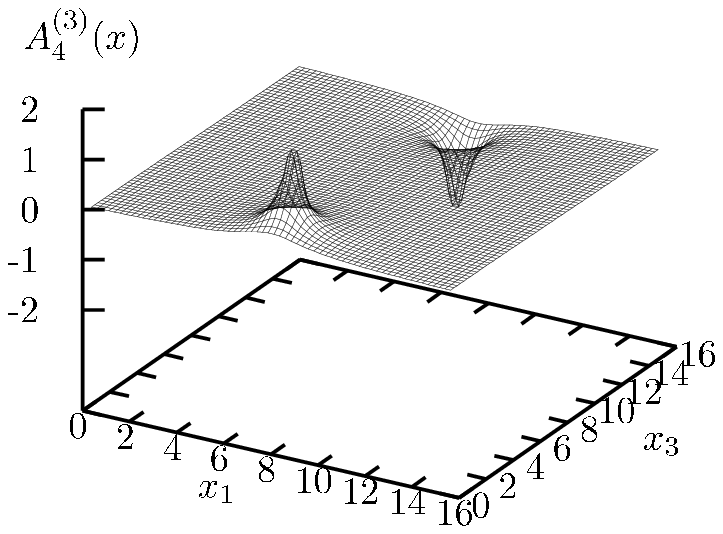}
\end{minipage}
}
\vs{2mm}
\caption{The three Lie algebra components of the Euclidean vector potential 
$A_\mu(x)$ in the algebraic gauge are shown for a dissociated caloron with 
maximally non-trivial holonomy $|\vec \omega|=0.25$, $\rho=2$ and $\beta=1$. 
The monopoles are located on the $x_3$-axis.}
\label{fig:fig4B}
\end{figure}

We begin with studying the structure of the Dirac string gauge field for a 
single, well-dissociated caloron, \ie $\rho\gg\beta$, in the algebraic gauge 
and with maximally non-trivial holonomy $\omega = 0.25$. For convenience we fix 
$\vec{\omega} \vec{\tau}= \omega\tau_3$ and  align the orientation of the 
monopole axis along the $x_3$-axis with the caloron center position located 
at the origin. We now consider the three Lie algebra components of 
the caloron vector potential
\beq
A_\mu(\tilde x)=A_\mu^{(a)}(\tilde x) \, \tau_a
\eeq
at positions $\tilde x=(\vec x,t)$ close to the connecting axis and 
between but not too close to the two monopole positions. In the large 
$\rho$ limit one easily derives from the analytical equations of
the caloron gauge fields (given in \cite{Kraan:1998kp,Kraan:1998pm}), 
that the vector potential $A_\mu(\tilde x)$ is dominantly Abelian and 
governed by the $A_{1,2}^{(3)}(\tilde x)$ components 
according to
\bea
\label{eq:DScond1}
\left|A_\mu^{(1,2)}(\tilde x)\right| &\ll& 
\left|A_{1,2}^{(3)}(\tilde x)\right|, \\
\left|A_{3,4}^{(3)}(\tilde x)\right| &\ll& 
\left|A_{1,2}^{(3)}(\tilde x)\right|. 
\label{eq:DScond2}
\eea
In fact, one can show that the $A_{1,2}^{(3)}(\tilde x)$ components are 
proportional to $1/(\lambda+\epsilon)$ in the large $\rho$ limit, where 
$\lambda\equiv \sqrt{x_1^2+x_2^2}$ is the distance to the monopole axis 
and $\epsilon\equiv\epsilon(\tilde x,\rho)$ is a small number, exponentially 
suppressed by the $\rho$ parameter. 
This is what we referred to with the term ``strongly peaking, but regular'' 
in the preceding sections. Furthermore, the gauge fields become
static in time and in the region between the monopoles also independent 
of the $x_3$-coordinate
\bea
\label{eq:DScond3}
A_{1,2}^{(3)}(\tilde x)&\approx& A_{1,2}^{(3)}(x_1,x_2,x_3,t=0) \; , \\
A_{1,2}^{(3)}(\tilde x)&\approx& A_{1,2}^{(3)}(x_1,x_2,x_3=0,t) \; .
\label{eq:DScond4}
\eea
These observations are illustrated in Figure~\ref{fig:fig4B} showing for 
the situation under discussion the behavior of the three Lie algebra 
components of the corresponding caloron vector potential in the 
$x_1$-$x_3$-plane. 

We will now find a gauge transformation that puts the dominant components 
$A_{1,2}^{(3)}(\tilde x)$ between the constituents to almost zero. From the 
above observations we conclude that the transformation
\beq
A_{\mu}^{'}(x) = e^{iG(x)\tau_3} \cdot A_\mu(x) \cdot e^{-iG(x)\tau_3} - 
\partial_\mu G(x) \tau_3  \; ,
\label{eq:xxx4}
\eeq
which we refer to as {\it DS transformation}, with a function 
$G(x)\equiv \tilde G(x_1,x_2)$ depending only on the $x_1$- and $x_2$-coordinate, 
will do the job provided that $\tilde G(x_1,x_2)$ fulfills the two conditions
\bea
\label{eq:eq4A}
\partial_1\tilde G{(x_1,x_2)} &=& A^{(3)}_1(x_1,x_2,x_3=0,t=0) \; , \\
\partial_2\tilde G{(x_1,x_2)} &=& A^{(3)}_2(x_1,x_2,x_3=0,t=0) \; .
\label{eq:eq4B}
\eea
This guarantees that at $x_3=0$ the formerly dominating $A_{1,2}^{(3)}$ 
components are completely erased. However, the existence of such a function 
$\tilde G$ is non-trivial. The above requirements Eqs.~(\ref{eq:eq4A}) and 
(\ref{eq:eq4B}) have to be consistent with the condition of integrability 
$\partial_1\partial_2\tilde G(x_1,x_2) \equiv \partial_2\partial_1\tilde G(x_1,x_2)$. 
A first indication of the existence of the desired function $\tilde G$ can be obtained 
from a simple consideration taking the vanishing action density of the Dirac string 
and its Abelian dominance into account.

However, an explicit calculation of the vector potentials 
$A_{1,2}^{(3)}(\tilde x)$ at $x_3=0$ and $t=0$ in the large $\rho$ 
limit for $\lambda \equiv \sqrt{x_1^2+x_2^2} \ll 
r\equiv |\vec x - \vec z_1| = s \equiv |\vec x - \vec z_2|$, 
where $\vec z_1$ and $\vec z_2$ denote the monopole positions, 
clarifies this question. It yields
\bea
\label{eq:DiracStringGaugeField1}
A^{(3)}_1(x_1,x_2,x_3=0,t=0) &=& 
- \frac{x_2}{\lambda} \left(\frac{1}{\lambda} 
- \frac{1}{2}\frac{\lambda}{r^2} 
+ O\left(\frac{\lambda^3}{r^4}\right)  \right) \; , \\
A^{(3)}_2(x_1,x_2,x_3=0,t=0) &=& 
\frac{x_1}{\lambda} \left(\frac{1}{\lambda} - 
\frac{1}{2}\frac{\lambda}{r^2} + 
O\left(\frac{\lambda^3}{r^4}\right)  \right) \; .
\label{eq:DiracStringGaugeField2}
\eea
While the first summand is integrable as presumed, the second summand is not. 
However, it is suppressed compared to the first one for small distances 
$\lambda$ to the monopole axis.
Using polar coordinates $(x_1,x_2)\rightarrow(\lambda,\alpha)$ to parametrize the $x_1$-$x_2$ 
plane at $x_3=0$ the requested function $\tilde G$ can now be defined sufficiently 
close to the monopole axis by the path integral
\beq
\tilde G{(\lambda,\alpha)} \equiv \int\limits_{0}^{\alpha}d\alpha^{\prime}\, 
\left( \begin{array}{*{1}{c}} -\lambda\sin(\alpha^{\prime}) \\
\lambda\cos(\alpha^{\prime}) \end{array} \right)  
\cdot  
\left( \begin{array}{*{1}{c}} A^{(3)}_1(\lambda,\alpha^{\prime},x_3=0,t=0) \\
A^{(3)}_2(\lambda,\alpha^{\prime},x_3=0,t=0) \end{array} \right)  
\stackrel{\rho\rightarrow\infty}{=} \alpha + O\left(\frac{\lambda^2}{r^2} \right) \; .
\eeq
where adequate mappings between polar coordinates $(\lambda,\alpha)$ and 
Cartesian coordinates $(x_1,x_2)$ are implicit. 
One sees that this function $\tilde G$ is not continuous. 
Close to the axis the function value has a $2\pi$ discontinuity 
at $\alpha=0$. The expression $e^{i\tilde G\tau_3}$, however, is continuous
for sufficiently small $\lambda$. 
Hence, the Dirac string vector potential for $0<\lambda\ll r,s$ is 
almost pure gauge in the limit $\rho\rightarrow\infty$.

So far, we were only referring to the case of maximally non-trivial holonomy, \ie
$|\vec \omega|=0.25$. For general non-trivial holonomy $|\vec \omega|\neq 0.25$ the 
ratio of the 3-dimensional monopole radii is given by $\omega/\bar\omega$, such that 
for a fixed value of $\rho$ the constituents start to overlap, as soon as $\omega$ or $\bar\omega$, 
respectively, becomes too small. In that case the DS-superposition scheme, to be presented in the
following, is no longer reasonable. However, even for small values 
of $\omega$ or $\bar\omega$ the constituents become well-separated again, if the value of $\rho$ is
sufficiently increased. In an adequately chosen region between the two constituents
the vector potential then again obeys the relations (\ref{eq:DScond1}, \ref{eq:DScond2}, 
\ref{eq:DScond3}, \ref{eq:DScond4}). This can easily be seen from the so-called far-field limit
\bea
A_\mu^{ff}(x) = -\frac{1}{2}\hat\omega\vec\tau \bar\eta^3_{\mu\nu}\partial_\nu\ln\phi^{ff}(x)
&\mbox{with}&
\phi^{ff}(x) = 
\frac{|{\vec x}-{\vec z}_2| + ({\vec x}-{\vec z}_2) \cdot {\vec e}_3}
     {|{\vec x}-{\vec z}_1| + ({\vec x}-{\vec z}_1) \cdot {\vec e}_3}
\label{eq:farfieldlimit}
\eea
which describes the caloron gauge fields at sufficiently large distances away from the
monopole centers~\cite{Bruckmann:2002vy}. Here, the same assumptions and notations as in Section~\ref{sec:Solutions}
are used and $\hat\omega\equiv\vec\omega/|\vec\omega|$. The far-field is completely Abelian and in the 
algebraic gauge it is independent of $\omega$ and hence, one can easily derive from (\ref{eq:farfieldlimit})
the same typical structure of the Abelian part of the Dirac string that was given in 
(\ref{eq:DiracStringGaugeField1},\ref{eq:DiracStringGaugeField2}). The DS-superposition scheme,
to be introduced in the following, is therefore also applicable (with the same function $\tilde G(x_1,x_2)$)
for general non-trivial holonomy with the only restriction that for a given value of the parameter $\rho$ the 
region between the constituents where the vector potential behaves as described 
in (\ref{eq:DiracStringGaugeField1},\ref{eq:DiracStringGaugeField2})
is decreased the more the holonomy deviates from its maximally non-trivial setting.

\subsection{Application to caloron superpositions}
\label{subsec:Application}

We now introduce an arbitrary, but localized object $A_\mu^{Obj}(x)$ to be 
located on top of the Dirac string of the previously considered caloron 
$A_\mu^{Cal}(x)$, which is still assumed to be centered at the origin with 
the monopole axis aligned along the $x_3$-direction for convenience. 
The superposition scheme resulting in the improved gauge field 
$A_\mu^{DS}(x)$ for the superposition of both gauge fields is now worked 
out in three steps.

\begin{figure}[h]
\centering
\mbox{
\begin{minipage}{0.5\linewidth}
\includegraphics[angle=0,width=1.0\textwidth]{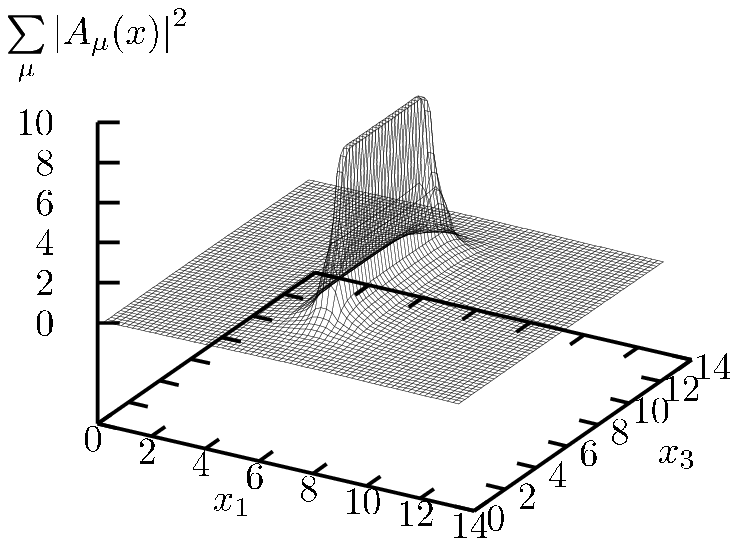}
\end{minipage}

\begin{minipage}{0.50\linewidth}
\includegraphics[angle=0,width=1.0\textwidth]{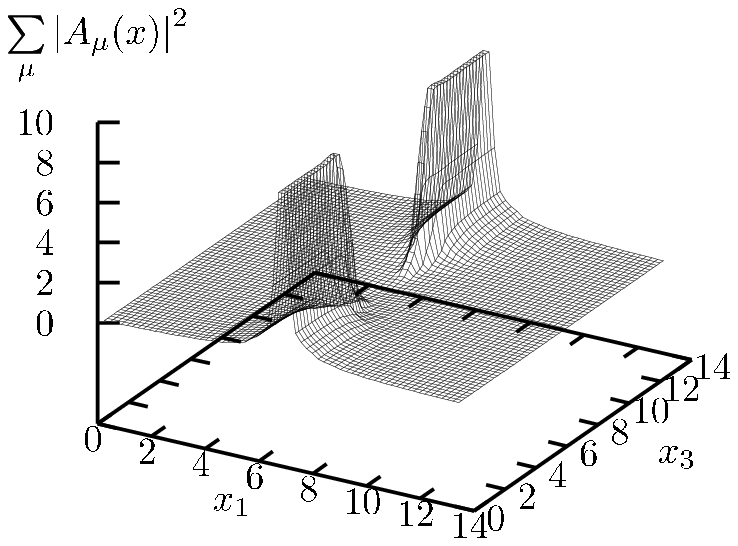}
\end{minipage}
}
\caption{The square norm $\sum_\mu|A_\mu(x)|^2$ of the vector potential 
in the algebraic gauge
before (left side) and after (right side) applying the DS gauge 
transformation Eq.~(\ref{eq:xxx4}) to the strongly dissociated caloron 
$A_\mu^{Cal}(x)$. The monopoles of the latter are located on the $x_3$-axis. 
By the transformation the gauge field in the space between the monopoles is 
approximately put to zero, while two new Dirac strings
are created outside the caloron again aligned parallel to the $x_3$-axis.}
\label{fig:fig4C}
\end{figure}

{\em (i)} The DS gauge transformation Eq.~(\ref{eq:xxx4}) is applied to 
$A_\mu^{Cal}(x)$ according to
\beq
A_\mu^{'Cal}(x) = e^{iG(x)\tau_3} \cdot A_\mu^{Cal}(x) \cdot e^{-iG(x)\tau_3} 
- \partial_\mu G(x) \tau_3 \; .
\label{eq:xxx5}
\eeq
where $G(x)\equiv\tilde G(x_1,x_2)$. 
This transformation reduces the vector potential of the Dirac string in the 
space between the two constituents. At the same time it creates two new 
Dirac strings of infinite length outside of the 
caloron which begin at the monopole locations as shown in 
Figure~\ref{fig:fig4C}.

{\em (ii)} The gauge fields, both brought to the algebraic gauge, are 
added resulting in the interim result $A_\mu^{'}(x)$ given as
\beq
A_\mu^{'}(x) = A_\mu^{'Cal}(x) + A_\mu^{Obj}(x) \; .
\label{eq:xxx6}
\eeq

{\em (iii)} The inverse gauge transformation is applied to 
$A_\mu^{'}(x)$ to remove the newly created Dirac strings and put them back 
in place, \ie back into the space between the two monopoles, yielding the 
improved gauge field $A_\mu^{DS}(x)$ of this {\it DS-superposition scheme} 
according to 
\bea
A_\mu^{DS}(x) &=& e^{-iG(x)\tau_3} \cdot A_\mu^{'}(x) \cdot e^{iG(x)\tau_3} 
+ \partial_\mu G(x) \tau_3 \nonumber \\ 
&=& \underbrace{A_\mu^{Cal}(x)}_{unchanged} 
+ \underbrace{e^{-iG(x)\tau_3} \cdot A_\mu^{Obj}(x) \cdot 
e^{iG(x)\tau_3}}_{rotated} \; ,
\label{eq:DSsuperposScheme}
\eea
where the appropriate gauge transformation to the periodic gauge is implicit.

To summarize, the three steps {\it (i) - (iii)} can all be combined together by
simply gauge rotating ({\it not} gauge transforming) the vector potential
$A_\mu^{Obj}(x)$ of the object to be located on top of the Dirac string before 
adding it to the unchanged field $A_\mu^{Cal}(x)$. 

Three further remarks are in order here. Firstly, this ansatz is also valid 
for anticalorons. Due to the transformation prescription 
Eq.~(\ref{eq:dualtoantidual}) for the construction of anticalorons out of 
calorons the appropriate function $G$ dealing with the Dirac string of the 
anticaloron even is the same. Secondly, this technique can be generalized 
to an arbitrary number of Dirac strings with arbitrary positions and spatial 
orientations. For this purpose it is sufficient to apply an adequate number
of such gauge rotations (one for each Dirac string to be dealt with) on the 
involved monopoles/calorons. And finally, the latter statement is true even if 
situations arise in which the two monopoles of a well separated caloron would need
to be gauge rotated differently. Such a situation arises, for example, in the
case of two intertwined calorons as sketched in Figure~\ref{fig:fig4F} for 
$\Delta h = 0$. In this case the two inner monopoles would need to be rotated
while the two outer ones {\it must not} be gauge rotated. However, we have seen that
the Dirac string connecting the two monopoles consists only of the $\vec\omega\vec\tau$
component. Since the needed gauge rotation is exactly around this axis in colour
space, the rotation has no effect on the Dirac string. Hence, the rotation can 
effectively be stopped at any point of the Dirac string, allowing the two
monopoles to be rotated independently of each other. Exactly this will be 
demonstrated in the next section.

\subsection {Comparing DS-improved solutions with the sum-ansatz}
\label{subsec:comparison}
The DS superposition scheme Eq.~(\ref{eq:DSsuperposScheme}) introduced above 
is now compared with the sum-ansatz~(\ref{eq:xxx3}). Both gauge fields 
$A_\mu^{DS}(x)$ and $A_\mu^{sum}(x)$, so far given in analytical
form, will be discretized on a lattice by a set of link variables according 
to Eq.~(\ref{eq:eq2Y}), thus making the gauge fields accessible to cooling 
procedures. The (in)stability of the improved and the unimproved 
configurations during the cooling process will give information about how 
close the obtained gauge fields are to classical configurations.
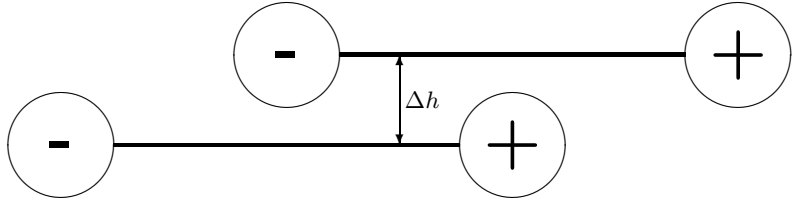
\begin{figure}[htb]
\centering
\setlength{\unitlength}{1.0cm}
\raisebox{0.0cm}{
\begin{picture}(12.0,3.0)(-6.0,-1.5)
   \put(-4.5,-0.6){\circle{1.4}}
   \put(+4.5,+0.6){\circle{1.4}}
   \put(-1.5,+0.6){\circle{1.4}}
   \put(+1.5,-0.6){\circle{1.4}}
   
   \linethickness{0.5mm}
   \put(-3.8,-0.6){\line(+1,0){4.6}}
   \put(3.8,+0.6){\line(-1,0){4.6}}
   
   \put(-4.5,-0.6){\makebox(0,0)[cc]{\textbf{\Huge -}}}
   \put(+4.5,+0.6){\makebox(0,0)[cc]{\textbf{\Huge +}}}
   \put(-1.5,+0.6){\makebox(0,0)[cc]{\textbf{\Huge -}}}
   \put(+1.5,-0.6){\makebox(0,0)[cc]{\textbf{\Huge +}}}
   
   \linethickness{0.05mm}
   \put(0.0,-0.6){\vector(0,+1){1.2}}
   \put(0.0,+0.6){\vector(0,-1){1.2}}
   \put(0.3,0){\makebox(0,0)[cc]{$\Delta h$}}
   
\end{picture}}
\caption{Sketch of the caloron composition to be used for the comparison 
of the superposition schemes. The two calorons are well-dissociated 
($\rho=1.5$, $\beta=1$), and carry maximally non-trivial holonomy
$|\vec \omega|=0.25$. The distance between their (parallel) monopole 
axes is denoted as $\Delta h$. We will consider the case $\Delta h=0$, 
where one obtains a pair of intertwined calorons penetrating 
each other with the Dirac string, and $\Delta h \ne 0$. 
The signs ($\pm$) correspond to the magnetic charge of the monopoles 
and the straight lines symbolize the Dirac strings.}
\label{fig:fig4F}
\end{figure}

For the comparison a simple composition made out of two {\it intertwined} 
calorons ($\Delta h=0$) as sketched in Figure~\ref{fig:fig4F} is used. 
Both calorons are well-dissociated ($\rho=1.5$, $\beta=1$), and each of them 
penetrates with its Dirac string one monopole of the other caloron. 
Certainly, one expects the selfduality to be completely destroyed at the 
inner monopole positions.

\begin{figure}[htbp]
\centering
\includegraphics[angle=270,width=1.0\textwidth]{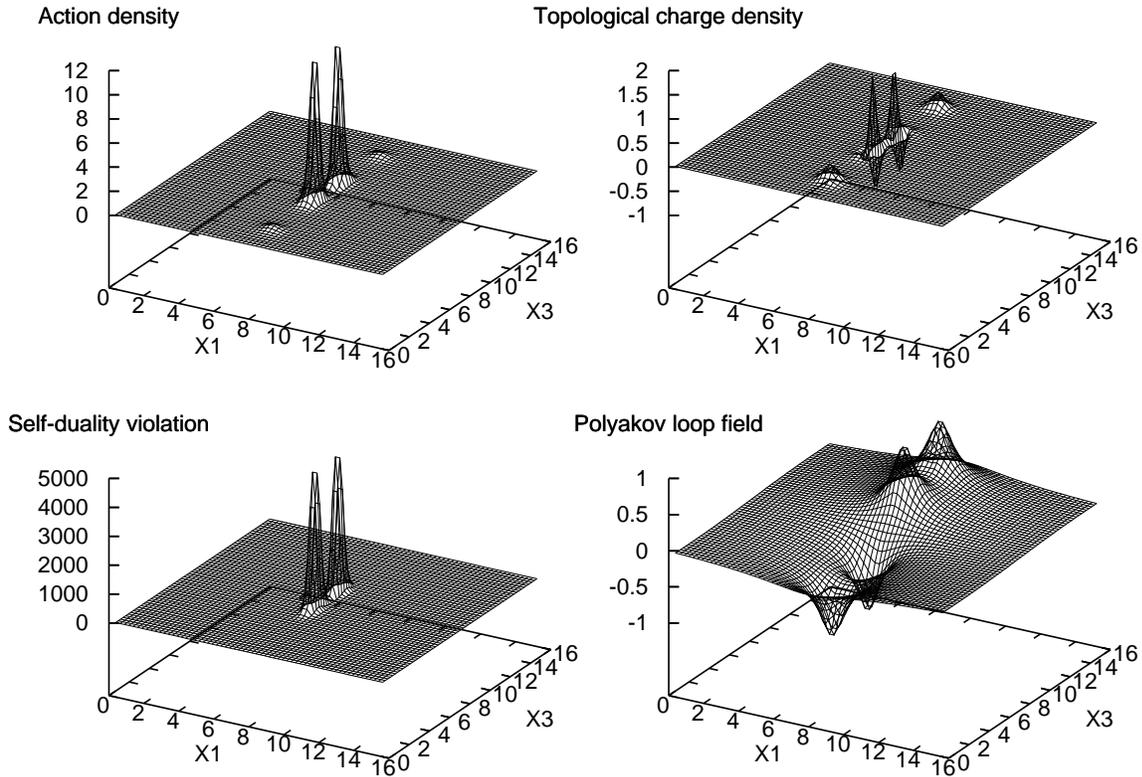}
\caption{Action density in instanton units, topological charge density, 
Polyakov loop field $\pol(\vec x)$ and the local violation of selfduality 
$\delta(x)$ for the gauge field configuration obtained by the  
sum-ansatz from the caloron composition with $\Delta h=0$ as sketched in 
Figure~\ref{fig:fig4F}.}
\label{fig:fig4D}
\end{figure}

Figure~\ref{fig:fig4D} shows the action density, the topological charge 
density, the Polyakov loop field $\pol(\vec x)$ and the local violation of 
selfduality $\delta(x)$, as defined in Eq.~(\ref{eq:eq4N}), for the 
configuration obtained from the sum-ansatz. The equation of motion is 
strongly violated at the locations of the inner monopoles as expected. 
This can be seen looking at the local violation of selfduality 
$\delta(x)$ as well as deduced from the fact that topological charge 
density and action density do not match at all in these regions. 
Moreover, the violation is strong enough to have strong impact also 
on the global measure $\gamma$, \ie the ratio of overall action and 
topological charge as defined in Eq.~(\ref{eq:DefOfGamma}), which 
should be unity for exactly selfdual configurations. 
From Table~\ref{tab:tab4A} we find $\gamma=5.7 \gg 1$. 

In order to reduce the violation of selfduality $\epsilon$-cooling 
with $\epsilon=0$ is applied to this gauge field configuration. 
Table~\ref{tab:tab4A} shows the corresponding cooling history, \ie 
the number of cooling steps together with the corresponding values 
for the action, the topological charge and $\gamma$. By the strong
change of the values for the total action and topological charge one sees 
that the cooling process leads to essential alterations of the configuration 
originally provided by the sum-ansatz.
\begin{table}[htbp]
\centering
\begin{tabular}{|c|c|c|c|}
\hline
Cooling steps &Action $S[A]$& Topological charge $Q[A]$ & $\gamma=S/Q$\\ 
\hline
0             &   9.209        &   1.612           &  5.714   \\ 
1             &   5.046        &   1.643           &  3.072   \\ 
5             &   2.710        &   1.836           &  1.476   \\ 
10            &   2.124        &   1.934           &  1.098   \\ 
20            &   1.972        &   1.935           &  1.020   \\ 
50            &   1.938        &   1.933           &  1.002   \\ 
100           &   1.934        &   1.933           &  1.001   \\ 
\hline
\end{tabular}
\caption{Cooling history of the gauge field configuration obtained 
by the sum-ansatz from the two-caloron composition with $\Delta h=0$ as 
sketched in Figure~\ref{fig:fig4F}. The total action $S[A]$, the total 
topological charge $Q[A]$ as well as the ratio $\gamma=S[A]/Q[A]$
are shown after certain numbers of $\epsilon$-cooling steps with 
$\epsilon=0$.}
\label{tab:tab4A}
\end{table}

\begin{figure}[htbp]
\centering
\includegraphics[angle=270,width=1.0\textwidth]{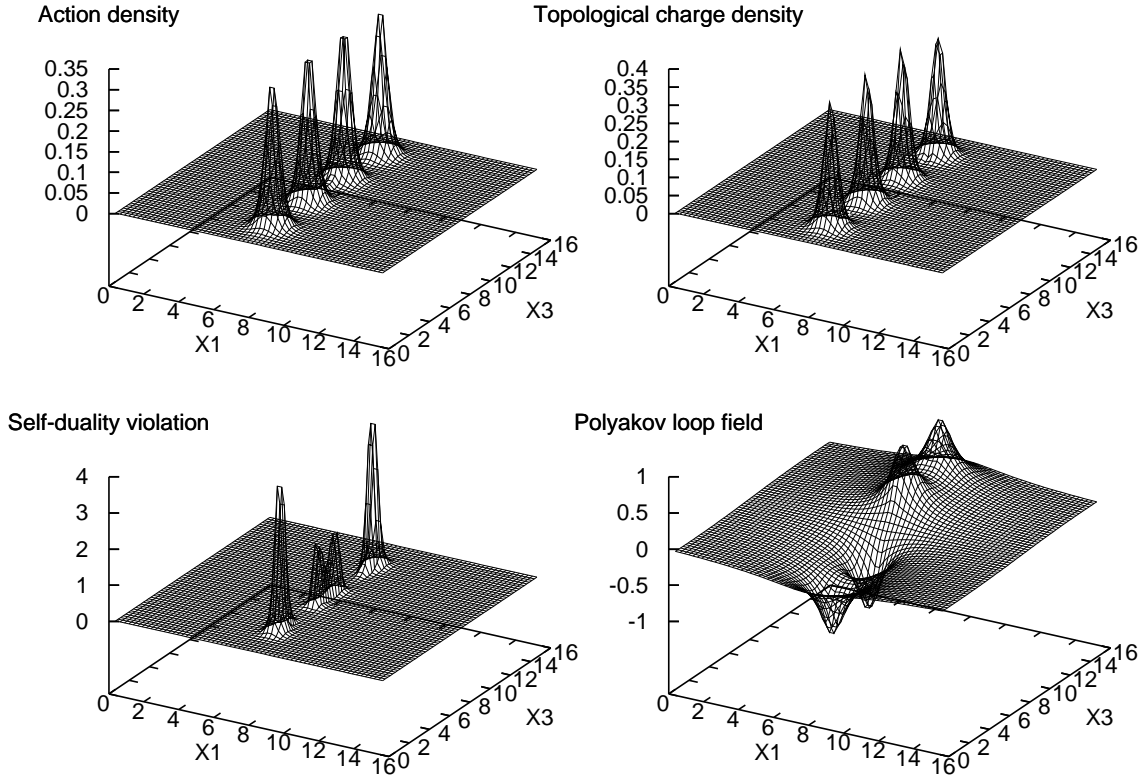}
\caption{The same as Figure~\ref{fig:fig4D} for the gauge field configuration 
obtained by the DS-improved superposition scheme from the caloron composition 
with $\Delta h=0$ as sketched in Figure~\ref{fig:fig4F}.}
\label{fig:fig4E}
\end{figure}

In Figure \ref{fig:fig4E} the same composition of calorons is shown but 
this time after the DS improvement has been applied. One sees that now 
the action density and the topological charge density do approximately 
match and also the local violation of selfduality $\delta(x)$ is 
by a factor $O(10^3)$
smaller compared to the unimproved case. Furthermore, we learn from 
Table~\ref{tab:tab4B} that the overall action surplus as obtained from the 
quantity $\gamma$ is approximately $6\,\%$,  thus much smaller 
than for the previously discussed sum-ansatz. Besides this observation 
we can conclude from the cooling history presented in Table~\ref{tab:tab4B} 
that the DS-improved configuration already is relatively close to a classical 
configuration, since there occurs only little alteration of the 
configuration during the fast convergence of the cooling process.

\begin{table}[htp]
\centering
\begin{tabular}{|c|c|c|c|}
\hline
Cooling steps & Action $S[A]$  & Topological charge $Q[A_\mu$] & $\gamma=S/Q$\\
\hline
0             &   2.073        &     1.941         &  1.068   \\ 
1             &   1.969        &     1.940         &  1.015   \\ 
5             &   1.940        &     1.937         &  1.001   \\ 
10            &   1.936        &     1.936         &  1.000   \\ 
20            &   1.934        &     1.934         &  1.000   \\ 
50            &   1.933        &     1.933         &  1.000   \\ 
100           &   1.931        &     1.932         &  0.999   \\ 
\hline
\end{tabular}
\caption{Cooling history with $\epsilon = 0$ for the gauge field configuration 
obtained by the DS-improved superposition scheme from the caloron composition 
with $\Delta h=0$ as sketched in Figure~\ref{fig:fig4F}, to be compared at 
corresponding step number with Table~\ref{tab:tab4A}.}
\label{tab:tab4B}
\end{table}

\begin{figure}[htb]
\centering
\mbox{
\begin{minipage}{0.5\linewidth}
\includegraphics[angle=0,width=1.0\textwidth]%
{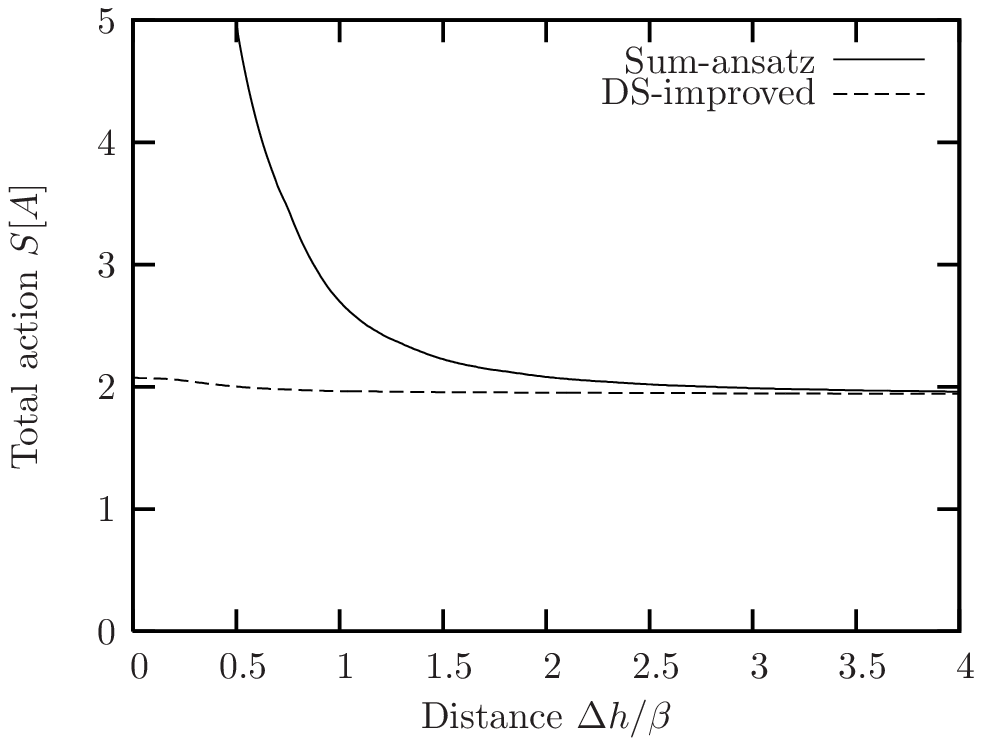}
\end{minipage}
\begin{minipage}{0.50\linewidth}
\includegraphics[angle=0,width=1.0\textwidth]%
{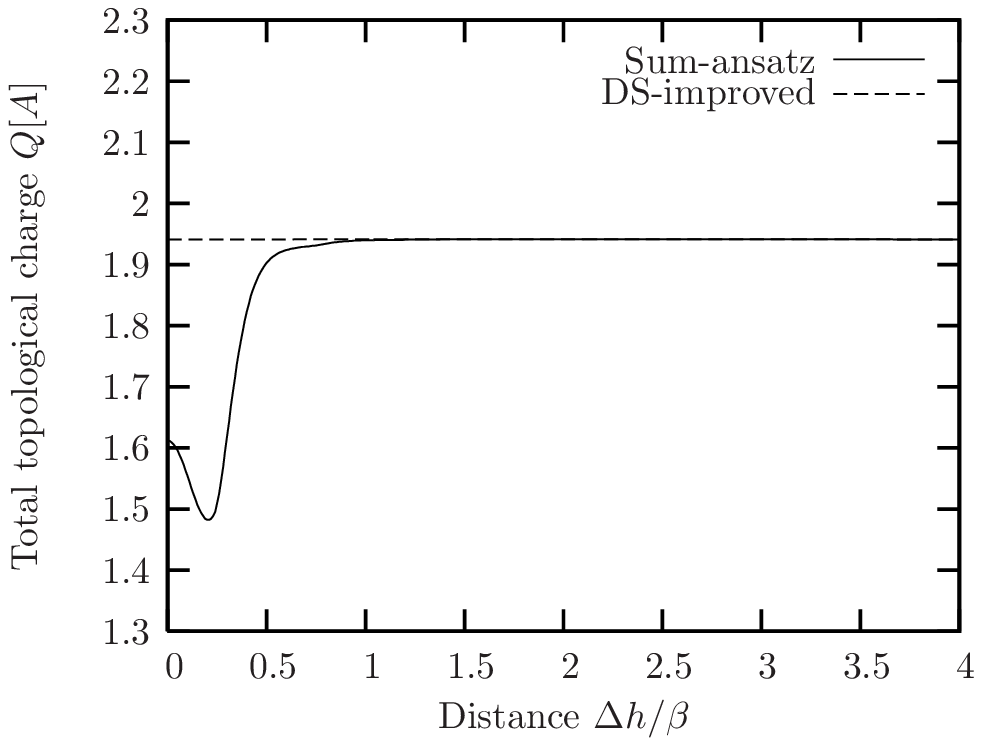}
\end{minipage}
}
\caption{Dependence of the total action $S[A]$ in instanton units $S_I$ (left) 
and of the total topological charge $Q[A]$ on the transverse distance 
$\Delta h$ between the monopole axes for the two-caloron composition 
sketched in Figure~\ref{fig:fig4F}. 
The results from the sum-ansatz (solid line) are compared with 
the results of the DS-improved superposition scheme (dashed line).}
\label{fig:fig4Jff}
\end{figure}

Finally, let us answer the question how much the benefit of the DS-improvement 
depends on the transverse separation $\Delta h$ between the monopole axes of 
the two calorons. For large separation $\Delta h \rightarrow \infty$ both
superposition schemes should be equivalent. In Figure~\ref{fig:fig4Jff} we 
show the dependence of the total action and topological charge on the 
separation $\Delta h$ for both superposition schemes. As expected, 
the sum-ansatz and the DS-improvement yield the same results for large 
separations $\Delta h$. For $\Delta h < 2\beta$, which corresponds to the 
case when Dirac string and monopole 
touch each other, since the 2D-radius of the first
and the 3D-radius of the latter are approximately $\beta$,
the sum-ansatz produces strong Dirac string-monopole interactions and thus
a great action surplus, while the DS-improved results are almost unaffected by
the transverse displacement. The latter observation also applies to the 
(lattice) topological charge of the improved configuration. The charge differs by 
$\approx 5\,\%$ from $Q=2$ even for asymptotic displacement which is a 
finite-volume or discretization effect. For the unimproved case the 
topological charge deviates much stronger from $Q=2$ for $\Delta h < \beta$,
which is due to the finite lattice discretization
and the vector potential becoming rather unsmooth in the interacting region.

We can conclude that the major part of the action surplus induced by 
caloron-Dirac string interactions can successfully be removed by the 
DS-superposition scheme, which becomes important at monopole-Dirac string 
separations of $\Delta h = 2\beta$. Alternatively, we have also seen that 
this action surplus can also be reduced by applying a moderate number of 
cooling steps. Due to the tremendous strength of the caloron-Dirac string 
interaction such a cooling procedure would definitely be a necessary step 
in the semiclassical modeling based on well-dissociated calorons superposed 
by the sum-ansatz. However, whereas the application of a moderate number
of cooling steps to a multi-caloron system seems to be acceptable, as long 
as the caloron or monopole radii are large compared to the lattice spacing, 
it becomes quite problematic for smaller calorons, since these small objects 
tend to fall through the grid during the cooling process. 
Cooling is similarly problematic for mixed caloron-anticaloron superpositions.
The DS-superposition scheme therefore provides a very interesting technique
for the construction of multi-caloron systems containing strongly dissociated 
calorons as well as small calorons, both of selfdual and anti-selfdual 
character.

\section{Improving selfduality by applying the ADHM formalism}
\label{sec:PseudoADHM}

The ADHM formalism~\cite{Atiyah:1978ri,Atiyah:1979iu} has been originally 
formulated by Atiyah, Hitchin, Drinfeld and Manin in 1978 to provide a 
proper framework for the construction of arbitrary (anti-)selfdual solutions 
of the Yang-Mills equation of motion in flat space-time, \ie for $T=0$. 
The same formalism, however, can also be applied to obtain classical gauge 
field configurations at finite temperature $T\neq 0$ and with arbitrary 
holonomy~\cite{Kraan:1998kp,Kraan:1998pm}. For the convenience of the reader
we will briefly review the basics of the ADHM formalism and then demonstrate, 
how the selfduality of the superposition of close calorons can be improved using 
this framework. Finally, we will again compare the improved solutions with the 
corresponding results of the naive sum-ansatz.

\subsection{Short summary of the ADHM formalism}
\label{subsec:summary}

Formally, there are two construction frameworks, the $Sp(N)$ ADHM formalism 
based on quaternions and the $SU(N)$ ADHM formalism based on complex spinors, 
and since $SU(2) \equiv Sp(1)$ both are applicable for $SU(2)$ instantons 
or calorons. However, it could be shown, at least for topological charge 
$Q=1$, that both approaches are identical~\cite{Atiyah:1978ri,Atiyah:1979iu}. 
Here, we will restrict ourselves to the $Sp(1)$ construction scheme for 
charge $Q = k > 0$ instantons or calorons. Anti-selfdual gauge fields can then be 
obtained from the corresponding selfdual gauge fields by 
Eq.~(\ref{eq:dualtoantidual}).

The basic ingredients of the $Sp(1)$ ADHM formalism for the construction of 
a charge $k \in \N$ instanton are a $k\times k$ symmetric, invertible, 
quaternionic matrix $B$ and a quaternionic row vector $\lambda$.
Using the Pauli matrices to define $\sigma_i \equiv -i\tau_i$ the space-time position 
$x=(\vec x,t)$ is encoded as a quaternion $x= x_i \sigma_i + t \ID$
allowing to define the space-time position dependent matrix $B(x)$ by
\beq
B{(x)} = B - x \cdot \ID_{k,k} \; ,
\label{eq:xxx12}
\eeq
where $\ID_{k,k}$ is the $k\times k$ quaternionic unit matrix. 
The vector potential $A_\mu(x)$ corresponding to the input data $B$ and 
$\lambda$ of this construction scheme is then obtained from the
quaternionic solution vector $u(x)$ of the following linear system of equations 
\beq
B^\dagger{(x)} \cdot u{(x)} = \lambda^\dagger \; ,
\label{eq:eq4J}
\eeq
which has to be solved for every point $x$ in space-time. 
The vector potential is then given by the simple equation
\beq
iA_\mu(x) = w^\dagger{(x)} \partial_\mu w{(x)} \; ,
\label{eq:eq4C}
\eeq
where a projection to the hyperspace spanned by 
$\{\sigma_1,\sigma_2,\sigma_3\}$ is implicit and $w(x)$ is just the 
solution vector $u(x)$ scaled by the potential $\phi(x)$ according to
\bea
\label{eq:eq4L}
w{(x)} &=& \sqrt{\frac{1}{\phi}} \cdot u{(x)} \; , \\
\phi{(x)} &=& 1 + u^\dagger{(x)}\cdot u{(x)} \in \re \; .
\label{eq:eq4K}
\eea
The vector potentials obtained from Eq.~(\ref{eq:eq4C}) solve the 
equation of motion in flat $\re^4$ and yield the topological charge 
$k$ if and only if the {\it ADHM constraint}
\bea
\label{eq:eq4D}
&\Delta^\dagger{(x)} \cdot \Delta{(x)} \in Mat(k,k,\re)& \; , \\
&\Delta{(x)} = 
\left(
\begin{array}{*{1}{c}} 
\lambda \\
B(x)  
\end{array}
\right)&
\label{eq:xxx13}
\eea
is fulfilled, where $Mat(k,k,\re)$ denotes the set of {\it real-valued} 
$k\times k$ matrices. In that case the $k$ eigenvalues of $B$ can be 
regarded as the positions of $k$ instantons in that solution.

One can easily see that this constraint is always fulfilled for $k=1$. 
Therefore, all initial parameters, \ie $B_{11}$ and $\lambda_1$, 
can be chosen freely, resulting in 8 free parameters for a selfdual 
solution in flat $\re^4$ with topological charge $k=1$. This is the 
single instanton solution with $B_{11}$ defining its position and 
$\lambda_1$ describing its size and global $SU(2)$ rotation.

The time-periodic caloron, defined on $\re^3 \times S^1$, 
with topological charge $Q=1$ and arbitrary holonomy 
can also be constructed by means of the ADHM 
formalism~\cite{Kraan:1998kp,Kraan:1998pm}. Applying the inverse of the
gauge transformation (\ref{eq:eq30F}) 
transforms the time-periodic caloron into a non-periodic, selfdual 
solution in flat $\re^4$ fulfilling the periodicity condition
\beq
A^{alg}_\mu(x+\beta\hat e_t) =  \polInf \cdot A^{alg}_\mu(x) \cdot 
\polInf^\dagger \; .
\label{eq:eq4H}
\eeq
It is the latter which can be constructed by 
the ADHM formalism. 

For the full parameter space of a single caloron the ADHM data $B_{mn}$ 
and $\lambda_m$, where $m,n \in \Z$, have been elaborated by Kraan and van 
Baal~\cite{Kraan:1998kp,Kraan:1998pm}. The corresponding ADHM data are
\beq
B_{mn} = (m+ \xi)\cdot \delta_{m,n} + 
\zeta^\dagger\hat{\omega}\vec{\sigma}\zeta \cdot 
\frac{\sin\left(2\pi(m-n)\omega\right)}{m-n}\cdot (1-\delta_{m,n})
\label{eq:eq4F}
\eeq
and
\beq
\lambda_m = e^{2\pi i m \vec{\omega}\vec{\tau}} \zeta \; ,
\label{eq:eq4G}
\eeq
where $\zeta, \xi$ are quaternions and $\beta=1$ for convenience. The 8 free 
parameters of a charge $Q=1$ caloron are encoded in $B_{mn}$ and $\lambda_m$. 
The quaternion $\xi$ describes the 4D space-time position of the caloron. 
The set of eigenvalues of $B$ is $\{\xi+j\beta : j\in \Z\}$ and corresponds
to a chain of instantons.  The size parameter $\rho$, the spatial orientation 
of the monopole axis, and the global $SU(2)$ rotation parameter are 
determined by
\beq
\zeta = \rho \cdot q, \quad q \in SU(2), \quad 
\hat{a}\vec{\tau} = q^\dagger\hat{\omega}\vec{\tau}q \; ,
\label{eq:xxx15}
\eeq
where $\hat{a}$ is the direction unit vector in three dimensions between 
the two caloron constituents.

This formalism can also be adopted for completely anti-selfdual 
instanton or caloron gauge fields, which can be related to their selfdual 
counterparts by Eq.~(\ref{eq:dualtoantidual}). The latter 
transformation can be built into the formalism by replacing 
Eq.~(\ref{eq:xxx12}) with 
\beq
B^{anti}{(x)} = B - x^\dagger \cdot \ID_{k,k} \; .
\label{eq:AntiCalxxx12}
\eeq
In that case one has 
\bea
u^{anti}(x)=u(x^\dagger),\quad & \phi^{anti}(x)=\phi(x^\dagger),\quad 
\mbox{and}\quad& w^{anti}(x)=w(x^\dagger)
\label{eq:uAnti}
\eea
leading to an exactly anti-selfdual gauge field. Its action density matches 
that of the corresponding selfdual counterpart if one inverts all spatial 
coordinates, \ie $x\leftrightarrow x^\dagger$. In particular, the caloron 
center $\xi$ becomes $\xi^\dagger$ for the anticaloron.

\subsection{Approximate multi-caloron solutions from ADHM formalism}
\label{subsec:approximate}

The ADHM formalism as described in the previous Section can in principle 
also be applied to multi-caloron configurations (all selfdual) with higher 
topological charge $Q > 1$. Thus, for a given set of $Q$ desired caloron 
center positions $\xi_i$ ($i=1,...,Q$) one is left with the problem of 
{\em (i)}~solving the non-linear ADHM constraint Eq.~(\ref{eq:eq4D}),
{\em (ii)}~fulfilling the periodicity condition Eq.~(\ref{eq:eq4H}), and 
{\em (iii)}~fixing the eigenvalues of $B$ according to a set of
$Q$ `instanton chains', \ie 
$\{\mbox{eigenvalues of }B\}=\{\xi_i+j\beta\hat e_t:i=1,...,Q;\,j\in \Z\}$.
For some special cases, e.g. all constituent positions located along one axis, 
corresponding solutions for the ADHM data have been 
presented~\cite{Bruckmann:2002vy,Bruckmann:2004nu}. 
But in all generality the ADHM constraint could not be solved due to its 
non-linear nature. Therefore, our idea is to construct approximate,
multi-caloron ADHM data $B^{\multi}$, $\lambda^{\multi}$ by using the 
well-known single caloron ADHM data $B^{(i)}$, $\lambda^{(i)}$
as building blocks. Thereby, we keep insisting on the periodicity condition 
{\em {ii)} and on the correct localization
of the caloron positions {\em {iii)} in the sense given above, while we 
allow for some violation of the ADHM constraint. However, since the ADHM 
data $B^{(i)}$, $\lambda^{(i)}$ of each single caloron solve the ADHM 
constraint exactly, the hope is that the newly composed multi-caloron ADHM 
data $B^{\multi}$, $\lambda^{\multi}$ can serve as an acceptable 
approximation for the true multi-caloron system. Later we will see that 
this {\it pseudo-ADHM ansatz} at least leads to gauge field configurations 
with higher selfduality than one would have obtained from the naive 
superposition of caloron gauge fields.

Here, we obtain the pseudo-ADHM data by simply merging the ADHM data of 
$Q$ single calorons according to
\bea
B^{\multi}_{mQ+i,nQ+j} &=& \Bigg\{   
\begin{array}{*{3}{ccl}}
B^{(i)}_{m,n} &:& \mbox{if }\; i=j\\
0             &:& \mbox{else}
\end{array} 
\quad
\mbox{with}\quad m,n\in \Z,\quad i,j=1,...,Q \; ,\\
\lambda^{\multi}_{mQ+i} &=& \lambda^{(i)}_{m}\quad 
\mbox{with} \quad m\in \Z, \quad i=1,...,Q \; .
\eea
This arrangement of the components of the single ADHM matrices 
$B_{m,n}^{(i)}$ guarantees the periodicity. Since the set of eigenvalues 
of $B^{\multi}$ is exactly the union of the eigenvalues of all $B^{(i)}$, 
the caloron positions are unchanged by this ansatz. Additionally, the ADHM
constraint still holds for every pair of columns of $\Delta(x)$ which 
have been filled with the ADHM data of the same caloron. On the other hand, 
for two columns belonging to different data sets $i,j\in \{1,...,Q\}$ 
($i\neq j$) one has
\beq
\left(\Delta^{\dagger,\multi}(x) \cdot \Delta^{\multi}(x)\right)_{mQ+i,nQ+j} 
= \lambda^{\dagger,(i)}_m \lambda^{(j)}_n, \quad m,n\in\Z.
\eeq
Therefore, the extent to which the ADHM constraint is violated is 
determined only by the vectors $\lambda^{(i)}$. 

This ansatz yields {\it exact} multi-caloron solutions in the case of 
trivial holonomy ($\polInf=\ID$) and equal global $SU(2)$ rotations of all 
involved calorons, \ie if all $\zeta^{(i)}$ are chosen proportional to each 
other, since then one always has 
$\lambda^{\dagger,(i)}_m \lambda^{(j)}_n\in\re$. The sizes and 4D-positions 
can then even be chosen arbitrarily. For instantons this observation is 
well-known as 't Hooft ansatz (unpublished, see Ref.~\cite{Jackiw:1976fs}).
Here, it will serve as our motivation to apply this ansatz also in the 
general case, where the holonomy is non-trivial.

We will now see that an analytical expression for the vector potentials 
can be derived from this ansatz. Let $u^{(i)}(x)$ denote the solution of 
Eq.~(\ref{eq:eq4J}) of the $i^{th}$ caloron. 
The corresponding solution for $B^{\multi}$, $\lambda^{\multi}$ 
can then be composed as
\beq
u^{\multi}_{mQ+i}(x) = u^{(i)}_m(x),\quad m\in \Z,\quad i=1,...,Q \; .
\label{eq:xxx16}
\eeq
From Eq.~(\ref{eq:eq4K}) one derives
\bea
\phi^{\multi}(x) &=& 1 + u^{\dagger,\multi}(x) u^{\multi}(x) =
1+\sum_{i=1}^Q u^{\dagger,(i)}(x)u^{(i)}(x) \nonumber\\
&=& 1 + \sum_{i=1}^Q \left(\phi^{(i)}(x)-1   \right) \; .
\label{eq:MultiPhi}
\eea
The vector potentials corresponding to $B^{\multi}$, $\lambda^{\multi}$ 
are now obtained from Eqs.~(\ref{eq:eq4C}) and (\ref{eq:eq4L}) yielding
\bea
iA_\mu^{\multi}(x) &=& 
\sqrt{\frac{1}{\phi^{\multi}(x)}} u^{\dagger,\multi}(x) \partial_\mu
\left(\sqrt{\frac{1}{\phi^{\multi}(x)}} u^{\multi}(x)  \right) 
= \frac{i}{\phi^{\multi}(x)} 
\sum_{i=1}^Q u^{\dagger,(i)}(x)\partial_\mu u^{(i)}(x) \nonumber\\
&=& \frac{i}{\phi^{\multi}(x)} \sum_{i=1}^Q \phi^{(i)}(x) A_\mu^{(i)}(x) \; ,
\label{eq:eq4M}
\eea
where $A_\mu^{(i)}(x)$ denotes the vector potential of a single caloron.
Here, the other superscript indexing the Lie algebra components is implicit.
We will refer to Eq.~(\ref{eq:eq4M}) as the {\it pseudo-ADHM 
superposition scheme} in the following. 
In words, the final result of this superposition ansatz 
is a weighted sum of the single caloron gauge fields with the weighing 
factor $\phi^{(i)}(x)/\phi^{\multi}(x)$ analytically 
given by Eqs.~(\ref{eq:MultiPhi}) 
and the analytical expression for $\phi(x)$ which are given in \cite{Kraan:1998kp,Kraan:1998pm}.~\footnote{After 
completion of this paper we learned that the superposition scheme
(\ref{eq:eq4M}) has also been considered by C.~Houghton, P.~van Baal and 
F.~Bruckmann (unpublished, private communication).}.

Furthermore, for the real modeling of the Yang-Mills vacuum, we need the 
extension of this ansatz to compositions of calorons and anticalorons.
Certainly, such configurations can neither be selfdual nor anti-selfdual. 
Nevertheless, we try such an extension. We incorporate such compositions 
into this construction scheme by replacing Eq.~(\ref{eq:xxx12}) 
with 
\beq
B^{\multi}_{mQ+i,nQ+j}(x) = \Bigg\{   
\begin{array}{*{3}{ccl}}
B^{\multi}_{mQ+i,nQ+j}-x\delta_{m,n}\delta_{i,j}
&:& \mbox{if the object }\; i\; \mbox{ is a caloron}\\
B^{\multi}_{mQ+i,nQ+j}-x^\dagger\delta_{m,n}\delta_{i,j}   
&:& \mbox{if the object }\; i\; \mbox{ is an anticaloron}
\end{array} 
\; ,
\label{eq:ExpandedPseudoADHMBMat}
\eeq
where $m,n \in \Z$ and $i,j=1,...,Q$. For a selfdual caloron the 
single ADHM data $B^{(i)}$ are chosen by Eq.~(\ref{eq:eq4F})
with $\xi= \xi_i$, where $\xi_i$ encodes
the desired caloron center position. 
For an anticaloron, however, we set $\xi=\xi^\dagger_i$ 
in Eq.~(\ref{eq:eq4F}) in order to locate the anticaloron at $\xi_i$. 

The solution vector $u^{\multi}(x)$ of the linear Eqs.~(\ref{eq:eq4J}) 
can then again be composed out of the single caloron solution vectors 
$u^{\multi}(x)$. Using Eq.~(\ref{eq:uAnti}) we obtain
\beq
u^{\multi}_{mQ+i}(x) = \Bigg\{ 
\begin{array}{*{3}{ccl}}
u_m(\lambda_i,\xi_i,x) &:& \mbox{if the object }\,i\,\mbox{ is a caloron}\\
u_m(\lambda_i,\xi_i^\dagger,x^\dagger)  &:& \mbox{if the object }\,i\,
\mbox{ is an anticaloron}
\end{array} 
\; ,
\label{eq:xxx18}
\eeq
where $m \in \Z$ and $i=1,...,Q$.
Here $u(\lambda,\xi,x)$ is the solution vector of Eq.~(\ref{eq:eq4J}) 
for a single caloron with the center position encoded as $\xi$. 
Since one has $\phi^{(i)}(\xi_i+x)=\phi^{(i)}(\xi_i+x^\dagger)$ as can
be seen from the analytical expression of $\phi(x)$ given in \cite{Kraan:1998kp,Kraan:1998pm},
the potential $\phi^{\multi}(x)$ for the extended ansatz 
is again given by Eq.~(\ref{eq:MultiPhi}), where the functions $\phi^{(i)}$ 
always denote the caloron $\phi$-potential for the caloron center 
position $\xi_i$. Finally, we arrive at
\beq
A_\mu^{\multi}(x) = \frac{1}{\phi^{\multi}(x)} 
\sum_{i=1}^Q \phi^{(i)}(x) \times 
\Bigg\{ 
\begin{array}{*{3}{ccl}}
A_\mu^{(i)}(x)        &:& \mbox{if the object }\,i\,\mbox{ is a caloron}\\
A_\mu^{anti,(i)}(x)   &:& \mbox{if the object }\,i\,\mbox{ is an anticaloron}
\end{array}
\; .
\label{eq:ExpandADHM4M}
\eeq
This is in slightly modified form the result that was already obtained 
in Eq.~(\ref{eq:eq4M}). The only difference is the appearance of the 
anti-selfdual vector potentials $A_\mu^{anti,(i)}(x)$ corresponding to 
anticalorons at some center positions $\xi_i$. 

A few further remarks are in order here for the orientation of the reader. 
First of all, it should be noted that the pseudo-ADHM superposition scheme 
recovers the sum-ansatz (\ref{eq:xxx3}) far apart from all caloron centers, 
since  one has $\phi^{(i)}(x)\rightarrow 1$ for $|\xi_i-x|\rightarrow\infty$. 
Hence, all weighing factors become unity in that limit. Very close to the 
monopole centers, however, one has $\phi^{(i)}(x)\gg 1$ 
and the corresponding gauge field becomes the dominant contribution 
to the weighted sum. More precisely, the corresponding weighing factor 
becomes unity while all others are suppressed. Therefore, this ansatz 
leads to better selfduality within the caloron cores and recovers the 
sum-ansatz far away from the cores. Applying this ansatz to compositions of 
calorons and anticalorons enforces the existence of regions where the 
solution is either nearly selfdual or nearly anti-selfdual, together with 
regions where the field is far from being either selfdual or anti-selfdual.

Finally, we demonstrate that this ansatz is a generalization of 
Shuryak's ratio-ansatz~\cite{Shuryak:1984kp}, a technique to improve the 
superposition of an instanton $A_\mu^{(1)}(x)$ and an antiinstanton 
$A_\mu^{(2)}(x)$. Given such a pair with positions $\xi_{1,2}$ and 
instanton size parameters $\rho_{1,2}$ one directly finds 
\bea
B^{(1)}= \xi_1, & B^{(2)} =  \xi_2^\dagger, & |\lambda^{(1,2)}|=\rho_{1,2}, \\
u^{(1)}(x)    =\frac{(\xi_1-x)}        {|\xi_1-x|^2} \lambda_1^\dagger, & 
u^{anti(2)}(x)=\frac{(\xi_2-x)^\dagger}{|\xi_2-x|^2} \lambda_2^\dagger, &
\phi^{(1,2)}(x) = 1 + \frac{\rho^2_{1,2}}{|\xi_{1,2}-x|^2} \; .
\eea
The separate (anti)instanton gauge field~\cite{Belavin:1975fg,'tHooft:1976fv} 
is related to $\phi^{(1,2)}$ by
\bea
A^{(1)}_\mu(x) = 
\frac{\tau_a}{2}O_1^{ab}\bar\eta^b_{\mu\nu}\partial_\nu\log\phi^{(1)}(x) 
& \mbox{ and } & 
A^{(2)}_\mu(x) = 
\frac{\tau_a}{2}O_2^{ab}\eta^b_{\mu\nu}\partial_\nu\log\phi^{(2)}(x)
\eea
where $\bar\eta^a_{\mu\nu}$, $\eta^a_{\mu\nu}$ denote the 't Hooft tensors 
and $\tau_a$ are the Pauli matrices. The matrices $O_{1,2}$ describe the 
color orientation which was encoded in $\lambda^{(1,2)}$. Inserting this 
into the {\it pseudo-ADHM} superposition scheme~(\ref{eq:ExpandADHM4M}) 
one immediately arrives at
\beq
A^{\multi}_\mu(x) = 
-2\frac{\frac{\tau_a}{2}O_1^{ab}\bar\eta^b_{\mu\nu}\rho_1^2y_{1\nu}/|y_1|^4
+ \frac{\tau_a}{2}O_2^{ab}\eta^b_{\mu\nu}\rho_2^2y_{2\nu}/|y_2|^4}{1 + 
\rho_1^2/|y_1|^2 + \rho_2^2/|y_2|^2}
\eeq
where $y_{1,2}\equiv x-\xi_{1,2}$ denote the 4-dimensional, real-valued 
distance vectors from the (anti)instanton centers.
This result exactly recovers the ratio-ansatz~\cite{Shuryak:1984kp} for 
instanton-antiinstanton pairs.

\subsection {Comparing pseudo-ADHM-improved solutions with the sum-ansatz}
\label{ComparingImproSolAndCool}

The pseudo-ADHM superposition scheme~(\ref{eq:ExpandADHM4M}) is now compared 
with the sum-ansatz~(\ref{eq:xxx3}). 
Again, the continuum gauge fields $A_\mu^{\multi}(x)$ and $A_\mu^{sum}(x)$,
respectively, will be discretized on a lattice to make the constructed gauge 
fields accessible to cooling procedures. Both the improved as well as the 
unimproved configurations are then exposed to a number of cooling steps. 
From the cooled configurations we assess the strength of the alteration 
during the cooling process and in this way indirectly the quality of the 
applied superposition scheme.
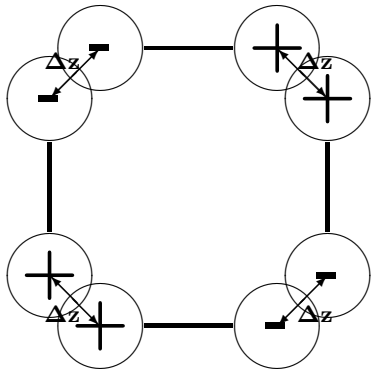
\begin{figure}[htp]
\centering
\setlength{\unitlength}{0.84cm}
\raisebox{0.0cm}{
\begin{picture}(8.0,6.0)(-4.0,-3.0)
   \put(-1.4,-2.2){\circle{1.4}}
   \put(-2.2,-1.4){\circle{1.4}}
   \put(1.4,2.2){\circle{1.4}}
   \put(2.2,1.4){\circle{1.4}}
   \put(1.4,-2.2){\circle{1.4}}
   \put(2.2,-1.4){\circle{1.4}}
   \put(-1.4,2.2){\circle{1.4}}
   \put(-2.2,1.4){\circle{1.4}}
   
   \linethickness{0.5mm}
   \put(-2.2,-0.7){\line(0,1){1.4}}
   \put(2.2,-0.7){\line(0,1){1.4}}
   \put(-0.7,-2.2){\line(1,0){1.4}}
   \put(-0.7,2.2){\line(1,0){1.4}}
   
   \put(-1.4,-2.2){\makebox(0,0)[cc]{\textbf{\Huge +}}}
   \put(-2.2,-1.4){\makebox(0,0)[cc]{\textbf{\Huge +}}}
   \put(1.4,2.2){\makebox(0,0)[cc]{\textbf{\Huge +}}}
   \put(2.2,1.4){\makebox(0,0)[cc]{\textbf{\Huge +}}}
   \put(1.4,-2.2){\makebox(0,0)[cc]{\textbf{\Huge -}}}
   \put(2.2,-1.4){\makebox(0,0)[cc]{\textbf{\Huge -}}}
   \put(-1.4,2.2){\makebox(0,0)[cc]{\textbf{\Huge -}}}
   \put(-2.2,1.4){\makebox(0,0)[cc]{\textbf{\Huge -}}}
   
   \linethickness{0.05mm}
   \put(-1.4,-2.2){\vector(-1,+1){0.78}}
   \put(-2.2,-1.4){\vector(+1,-1){0.78}}
   \put(+1.4,+2.2){\vector(+1,-1){0.78}}
   \put(+2.2,+1.4){\vector(-1,+1){0.78}}
   \put(+1.4,-2.2){\vector(+1,+1){0.78}}
   \put(+2.2,-1.4){\vector(-1,-1){0.78}}
   \put(-1.4,+2.2){\vector(-1,-1){0.78}}
   \put(-2.2,+1.4){\vector(+1,+1){0.78}}
   
   \put(-2.0,-2.0){\makebox(0,0)[cc]{$\mathbf{\Delta z}$}}
   \put(-2.0,+2.0){\makebox(0,0)[cc]{$\mathbf{\Delta z}$}}
   \put(+2.0,-2.0){\makebox(0,0)[cc]{$\mathbf{\Delta z}$}}
   \put(+2.0,+2.0){\makebox(0,0)[cc]{$\mathbf{\Delta z}$}}
\end{picture}}
\caption{Sketch of the caloron composition to be used for the comparison 
of the superposition schemes. The four calorons are well-dissociated 
($\rho=1.0$, $\beta=1$), and carry maximally non-trivial holonomy
$|\vec \omega|=0.25$. The signs ($\pm$) correspond to the magnetic 
charge of the monopoles and the straight lines symbolize the Dirac strings. 
The latter are aligned along the edges of a square.
We will consider the arrangement where the separation of two neighboring 
equal-charge monopoles is put equal to $\Delta z=0.28$, which is small 
compared to the 3-dimensional action lump size 
$\rho_{3D}\approx \beta=1$ of the undisturbed calorons.}
\label{fig:fig4I}
\end{figure}

For the purpose of comparison a simple planar composition of four calorons 
as sketched in Figure~\ref{fig:fig4I} is considered. 
The calorons are well-dissociated ($\rho=1.0$, $\beta=1$) and the monopole 
axes are aligned along the edges of a square such that 4 pairs of monopoles 
with equal magnetic charge are formed. The distance $\Delta z=0.28$ 
between two neighboring monopoles is small compared to the 3-dimensional 
radius $\rho_{3D}\approx\beta=1$ of their action lumps in the undisturbed case.
One therefore expects strong violations of selfduality at all monopole 
positions. The alert reader might have noticed that this example is 
constructed in such a way to avoid the problem (dealt with in 
Section \ref{sec:DiracString}) of interaction between the Dirac string 
of one caloron with the monopole constituents of another caloron.

Figure~\ref{fig:fig4G} shows the action density, the topological 
charge density, the Polyakov loop field $\pol(\vec x)$ and the local 
violation of selfduality $\delta(x)$ for the sum-ansatz. As expected, 
the equation of motion is strongly violated at the locations of the 
monopole pairs. This is shown by the local violation of selfduality 
$\delta(x)$ or by the fact that topological charge density and action 
density do not match in these regions. From the ratio $\gamma$ of 
action and topological charge taken from Table~\ref{tab:tab4C}
we furthermore see that the interaction of the close monopoles 
generates an action surplus of about $52\%$.
\begin{figure}[htb]
\centering
\includegraphics[angle=270,width=1.0\textwidth]{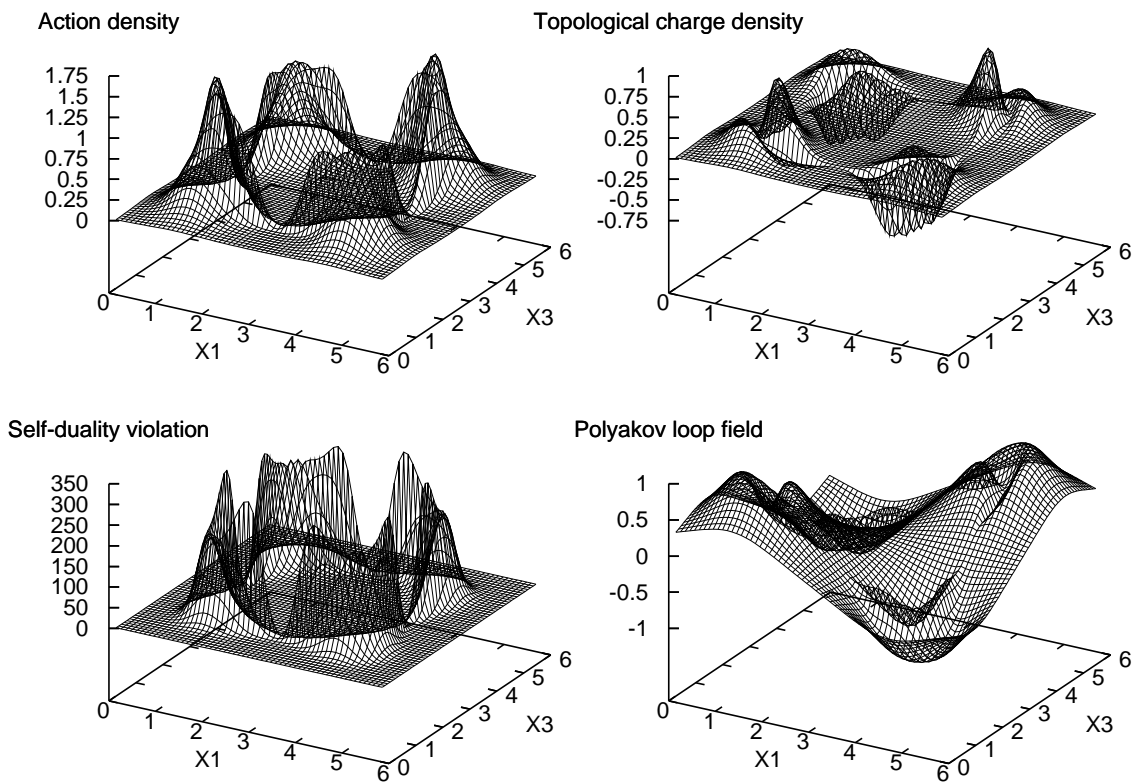}
\caption{Action density in instanton units, topological charge density, 
Polyakov loop field $\pol(\vec x)$ and the local violation of selfduality 
$\delta(x)$ for the gauge field configuration obtained by the  
sum-ansatz from the four-caloron composition with $\Delta z=0.28$ 
as sketched in Figure~\ref{fig:fig4I}.}
\label{fig:fig4G}
\end{figure}

In order to reduce the violation of selfduality $\epsilon$-cooling 
with $\epsilon=0$ is applied to this starting configuration. 
The corresponding cooling history is presented in Table~\ref{tab:tab4C}. 
One sees that a relatively small number $O(10)$ of cooling steps is 
sufficient to remove the action surplus. 
\begin{table}[htp]
\centering
\begin{tabular}{|c|c|c|c|}
\hline
Cooling steps &Action $S[A]$ & Topological charge $Q[A]$ & $\gamma=S/Q$\\ 
\hline
0             &   6.009     &    3.967             &  1.515 \\ 
1             &   5.148     &    3.980             &  1.293 \\ 
5             &   4.270     &    3.985             &  1.071 \\ 
10            &   4.090     &    3.985             &  1.026 \\ 
20            &   4.008     &    3.984             &  1.006 \\ 
50            &   3.984     &    3.984             &  1.000 \\ 
100           &   3.982     &    3.984             &  1.000 \\ 
\hline
\end{tabular}
\caption{Cooling history of the gauge field configuration obtained by 
the sum-ansatz from the caloron composition with $\Delta z=0.28$ as 
sketched in Figure~\ref{fig:fig4I}. The total action $S[A]$, the total 
topological charge $Q[A]$ as well as the ratio $\gamma=S[A]/Q[A]$
are shown after certain numbers of $\epsilon$-cooling steps with 
$\epsilon=0$.}
\label{tab:tab4C}
\end{table}

The same composition of calorons, this time generated by means of the 
pseudo-ADHM superposition scheme, is shown in Figure~\ref{fig:fig4H}. 
Clearly, the local violation of selfduality $\delta(x)$ is much smaller 
than for the previously discussed sum-ansatz. Furthermore, the 
topological charge and action density now resemble each other and 
the topological charge is now $Q=4$ with high precision. 
Besides that, the improved selfduality can also be recognized by the 
smaller action surplus, being approximately $21\,\%$ as obtained 
from the observable $\gamma$ in Table~\ref{tab:tab4D}. The corresponding 
cooling history is also presented in this Table. Three further remarks 
are in order here.

{\em (i)} It should be noted that the action lumps obtained from the 
pseudo-ADHM technique are much sharper localized than they are for 
an undisturbed caloron with the same parameters. This effect becomes 
even much stronger for further decreasing monopole distances $\Delta z$. 
The shrinkage of the monopoles is due to the weighing factor in 
Eq.~(\ref{eq:eq4M}) with which each single gauge field is 
multiplied. As discussed in the previous Section this weighing factor 
guarantees that close to the monopole centers the corresponding gauge 
field dominates the superposition such that the selfduality is improved. 

{\em (ii)} The pseudo-ADHM technique leaves the monopole center positions
unaltered. 

{\em (iii)} The outcome of the pseudo-ADHM technique can {\it not} be 
imitated by the sum-ansatz and some adequate adjustment of the parameters 
of the involved calorons. 

\begin{figure}[htb]
\centering
\includegraphics[angle=270,width=1.0\textwidth]{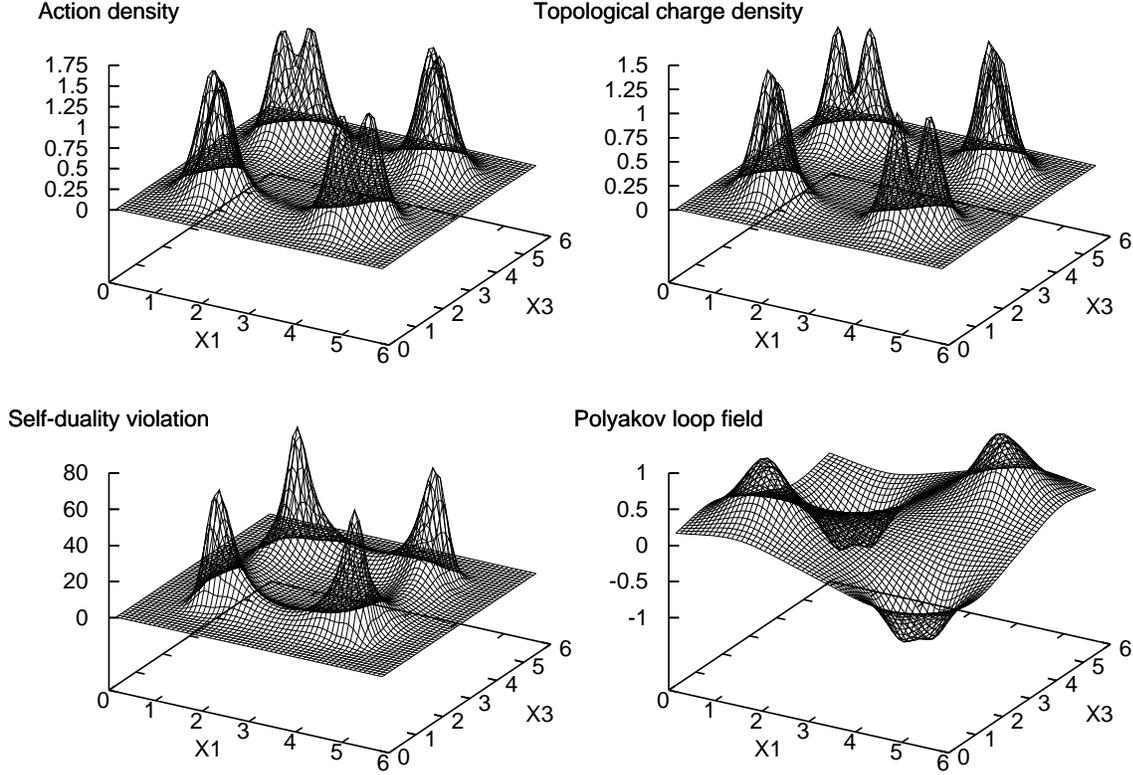}
\caption{Action density in instanton units, topological charge density, 
Polyakov loop field $\pol(\vec x)$ and the local deviation from selfduality 
$\delta(x)$ for the gauge field configuration obtained by the DS-improved 
superposition scheme from the caloron composition with $\Delta z=0.28$ 
as sketched in Figure~\ref{fig:fig4I}.}
\label{fig:fig4H}
\end{figure}

\begin{table}[htb]
\centering
\begin{tabular}{|c|c|c|c|}
\hline
Cooling steps &Action $S[A]$& Topological charge $Q[A]$ & $\gamma=S/Q$\\ 
\hline
0             &   4.856         &  3.999           &  1.214     \\ 
1             &   4.740         &  4.000           &  1.185     \\ 
5             &   4.422         &  4.000           &  1.105     \\ 
10            &   4.260         &  4.000           &  1.065     \\ 
20            &   4.122         &  3.999           &  1.031     \\ 
50            &   4.022         &  3.998           &  1.006     \\ 
100           &   4.001         &  3.997           &  1.001     \\ 
\hline
\end{tabular}
\caption{Cooling history of the gauge field configuration obtained by 
the DS-improved superposition scheme from the caloron composition with 
$\Delta z=0.28$ as  sketched in Figure~\ref{fig:fig4I}. The total 
action $S[A]$, the total topological charge $Q[A]$ as well as the 
ratio $\gamma=S[A]/Q[A]$ are shown after certain numbers of 
$\epsilon$-cooling steps with $\epsilon=0$.}
\label{tab:tab4D}
\end{table}

Next, it will be demonstrated that cooling alters the gauge 
field configuration obtained 
from the sum-ansatz strongly, whereas the improved gauge field given by the 
pseudo-ADHM approach is only modified gently thanks to its almost perfect
selfduality. Therefore, the action densities corresponding to both cases 
after applying 100 cooling steps are shown in Figure~\ref{fig:fig4J}. 
(The topological charge density is the same after cooling.) 
While the action density provided by the sum-ansatz has completely changed its 
appearance, the action density of the pseudo-ADHM improved configuration 
is only slightly altered. In the latter case the monopoles tend to drift apart
from each other, which implies that the monopole cores 
grow in their spatial extension and their maximal action density decreases.
One may therefore conclude, that the 
pseudo-ADHM improved gauge field is already close to a classical solution 
by construction, under the particular condition to consist of separable 
monopoles. For further improvement (cooling) it reacts by slightly 
separating the monopoles which by itself are stabilized by the weighted 
superposition.

\begin{figure}[htb]
\centering
\mbox{
\begin{minipage}{0.5\linewidth}
\includegraphics[angle=270,width=1.0\textwidth]%
{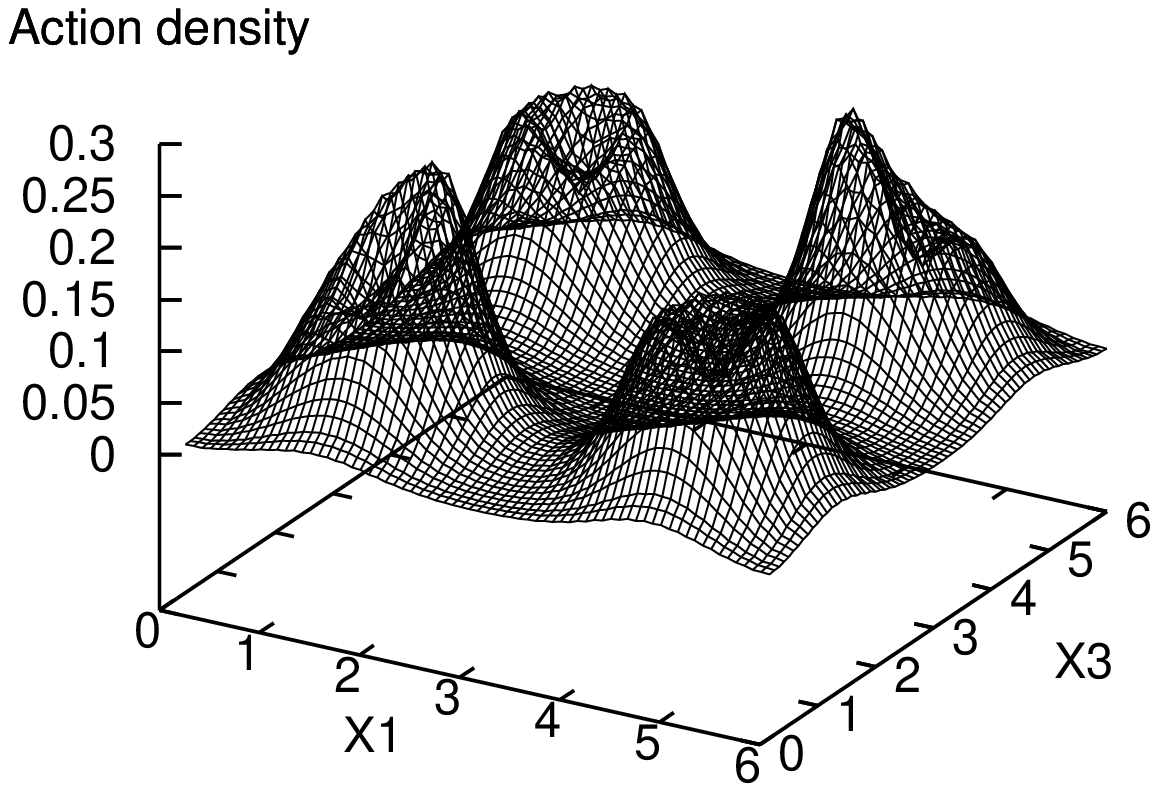}
\end{minipage}
\begin{minipage}{0.50\linewidth}
\includegraphics[angle=270,width=1.0\textwidth]%
{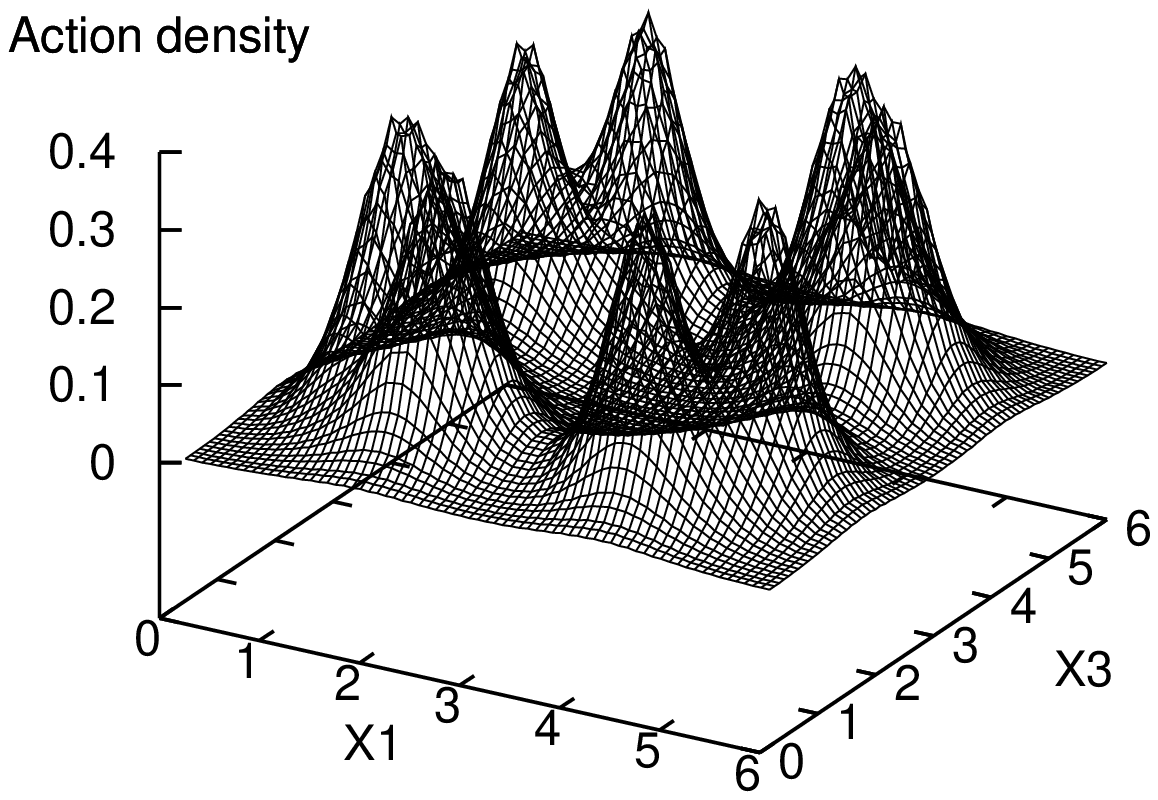}
\end{minipage}
}
\caption{Action densities in instanton units $S_I$ for the caloron 
composition sketched in Figure~\ref{fig:fig4I} after applying 100 
$\epsilon$-cooling steps with $\epsilon=0$. The gauge field configurations 
before cooling were constructed by the sum-ansatz~(left) and by the 
pseudo-ADHM technique~(right).                  }
\label{fig:fig4J}
\end{figure}

We should draw particular attention to the special volcano-like shape of 
the cooled configurations that have developed from the sum-ansatz. 
Such a structure (with local minima of action density in the center) is 
characteristic for $m=2$ monopoles and $m=-2$ antimonopoles and has also 
been observed in analytical solutions for charge $Q=2$ calorons with two equally
charged monopoles coming close to each other while the other two were sent
to infinity~\cite{Bruckmann:2004nu}.
In contrast, the cooled pseudo-ADHM improved configurations do not exhibit 
such kind of structure. Hence, it remains unclear whether the pseudo-ADHM 
improved solutions in our case can actually be regarded as 
being closer to the exact $Q=4$ solution in the configuration space, 
although their selfduality is certainly improved by far compared to the sum-ansatz.
\begin{figure}[htb]
\centering
\mbox{
\begin{minipage}{0.5\linewidth}
\includegraphics[angle=0,width=1.0\textwidth]%
{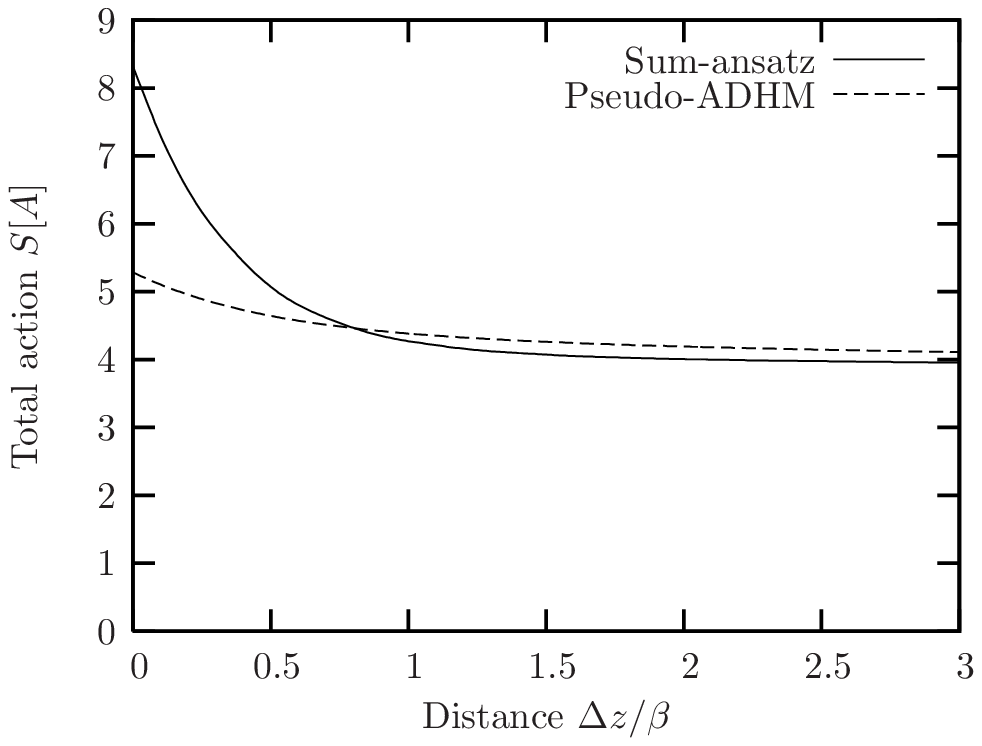}
\end{minipage}
\begin{minipage}{0.50\linewidth}
\includegraphics[angle=0,width=1.0\textwidth]%
{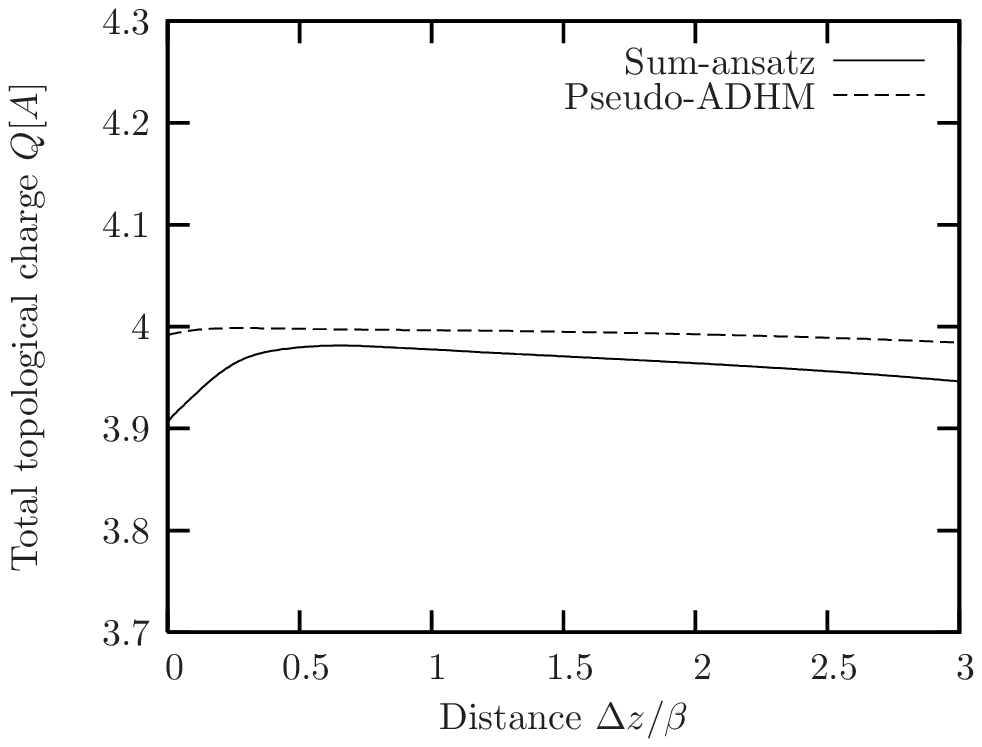}
\end{minipage}
}
\caption{Dependence of the total action $S[A]$ in instanton units 
$S_I$ (left) and of the total topological charge $Q[A]$ (right)
on the monopole separation $\Delta z$ for the caloron composition 
sketched in Figure~\ref{fig:fig4I}. The results from the sum-ansatz 
(solid line) are compared to the results of the pseudo-ADHM superposition 
scheme (dashed line).}
\label{fig:fig4Jgf}
\end{figure}

Finally, the dependence of the created action surplus on the monopole 
separation $\Delta z$ is of interest. Figure~\ref{fig:fig4Jgf}
shows the total action and topological charge as measured in the pseudo-ADHM 
improved and unimproved (sum-ansatz) configurations versus $\Delta z$.
Since the two superposition schemes become equivalent at large caloron 
separations as discussed in the previous Section,
the results for the total action converge to each other for 
$\Delta z\rightarrow\infty$. For $\Delta z/\beta < 0.8$ the action of 
the sum-ansatz exceeds the action of the improved construction. 
At closer separations the action surplus generated by the sum-ansatz 
becomes much larger than for the pseudo-ADHM improved configurations. The
improved superposition scheme hence becomes meaningful at 
small distances, here at $\Delta z<0.7\beta$. The topological charge 
is almost perfectly $Q=4$, independently of the distance $\Delta z$,
for the improved superposition, and relatively constant 
(within a few percent) for the sum-ansatz.

\subsection {Comparison with an exact $Q=2$ solution}
\label{ComparingExactQ2solution}

In the last section the gauge field configurations constructed by the pseudo-ADHM 
superposition scheme were shown to suffer much less from deviations from selfduality 
than those obtained by the sum-ansatz. Now we address the question whether these improved
configurations are moreover also capable of qualitatively imitating exact higher charge 
solutions for some appropriately chosen single-caloron parameter sets. 

For that purpose we selected an exact $Q=2$ solution which was provided to us by 
the authors of~\cite{Bruckmann:2004nu} and which is shown in Fig.~\ref{fig:figExactQ2solution}a,d,g.
We then fixed the optimal single-caloron parameter sets for both superposition
schemes independently by minimizing the square norm of the difference between the
action densities of the exact and the superposed configurations. The optimal 
configuration constructed by the pseudo-ADHM technique is shown in Fig.~\ref{fig:figExactQ2solution}b,e,h,
while Fig.~\ref{fig:figExactQ2solution}c,f,i shows the "optimal" configurations
obtained from the sum-ansatz.

Comparing Fig.~\ref{fig:figExactQ2solution}a, b, and c one clearly sees that it is indeed 
possible for the pseudo-ADHM ansatz to mimic the 4-monopole structure of the exact $Q=2$ solution 
despite their small separations, while this is not the case for the sum-ansatz. 
For the latter ansatz the solution is completely spoilt by strong Dirac string interactions,
which do not occur for the pseudo-ADHM technique. One should also notice that these
Dirac string interactions can not be suppressed by the DS superposition scheme due to the
weak caloron dissociation. Besides the strong improvements concerning the local self-duality
also the overall measure $\gamma=S[A]/Q[A]$ for self-duality is again obviously better for the 
pseudo-ADHM ansatz. For the unimproved configuration we find $\gamma = 1.221$ and 
$\gamma = 1.066$ for the improved configuration. Since the self-duality aspects 
have already been discussed in the previous section, they are not explicitly demonstrated in 
detail here again.

However, for the pseudo-ADHM ansatz it can clearly be observed in Fig.~\ref{fig:figExactQ2solution}e,h
that those monopoles which were initially enclosed in the same caloron are still stronger connected 
(see Fig.~\ref{fig:figExactQ2solution} h) than those initially belonging to different calorons 
(see Fig. \ref{fig:figExactQ2solution}e). This asymmetry is due to the fact
that there is no mechanism included within the pseudo-ADHM ansatz to unify the
single calorons into one single, higher charge configuration. Therefore, this scheme will always 
preserve the single-caloron character of the superposition. This is because the single ADHM-data
of the involved calorons were merged in the simplest possible manner. However it might be surprising 
to see how well this improved technique also works out with respect to imitating 
this exact $Q=2$ solution in contrast to the sum-ansatz.

\begin{figure}[htbp]
\centering
\mbox{
\begin{minipage}{0.33\linewidth}
\includegraphics[angle=0,width=1.0\textwidth]{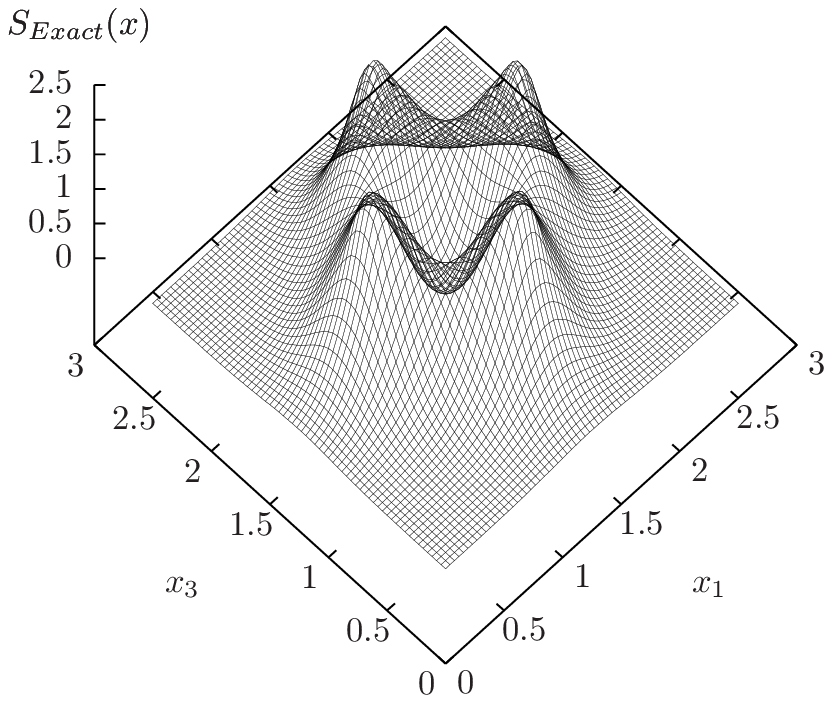}
a
\includegraphics[angle=0,width=1.0\textwidth]{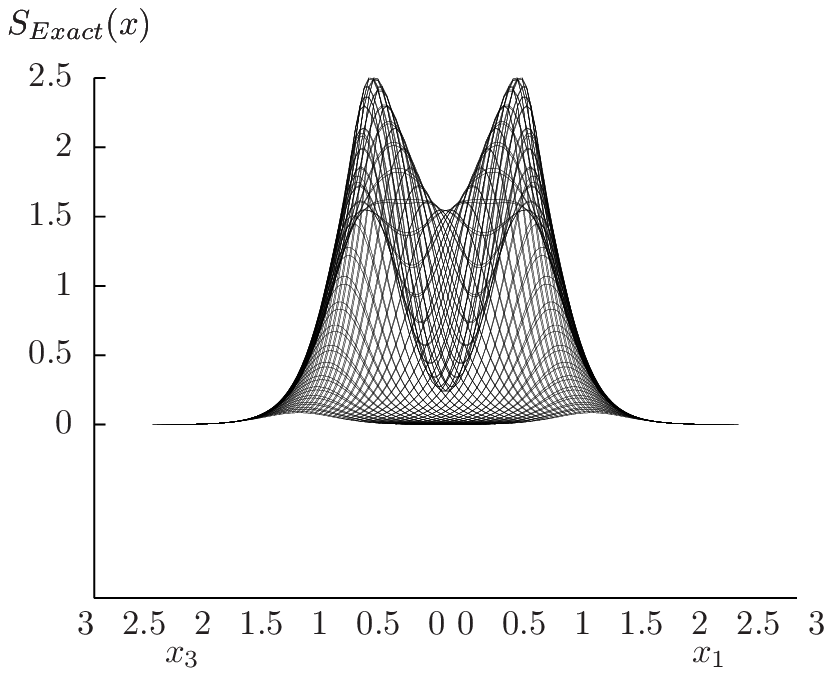}
d
\includegraphics[angle=0,width=1.0\textwidth]{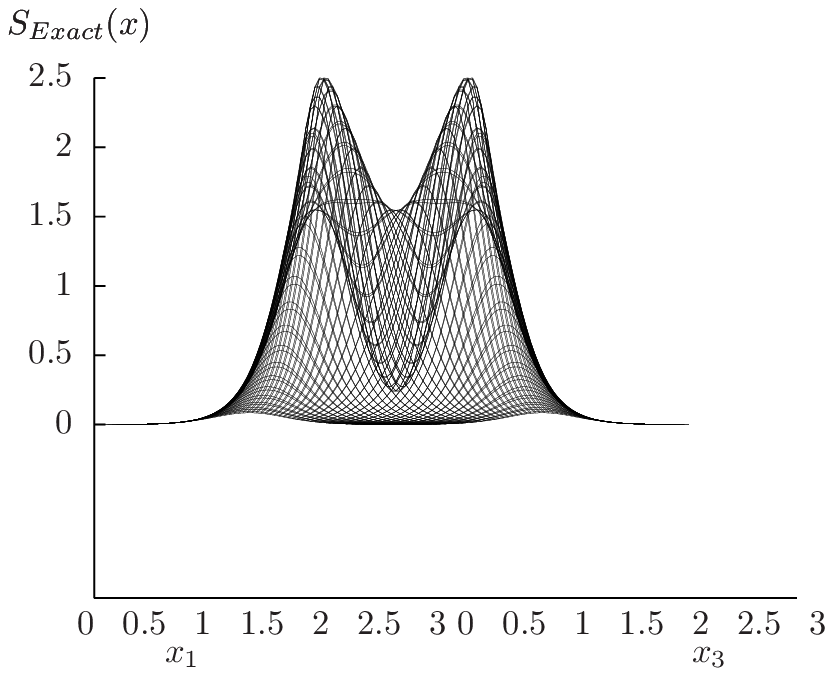}
g
\end{minipage}
\begin{minipage}{0.33\linewidth}
\includegraphics[angle=0,width=1.0\textwidth]{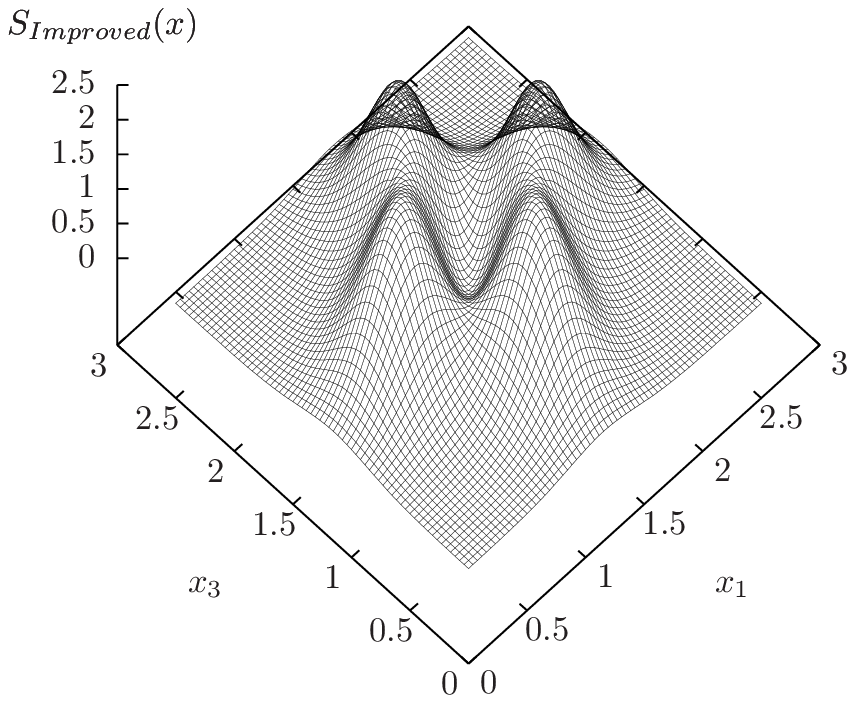}
b
\includegraphics[angle=0,width=1.0\textwidth]{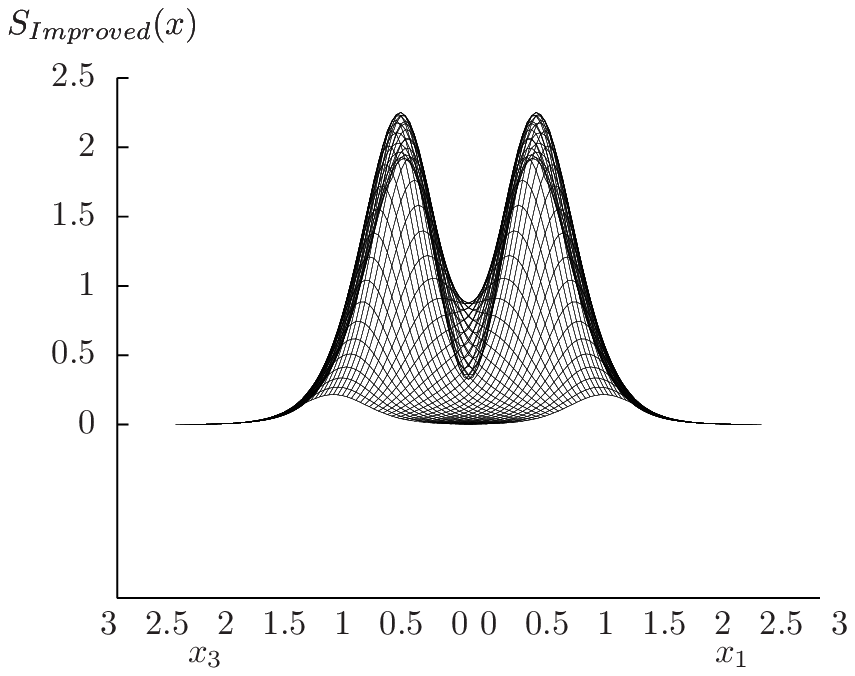}
e
\includegraphics[angle=0,width=1.0\textwidth]{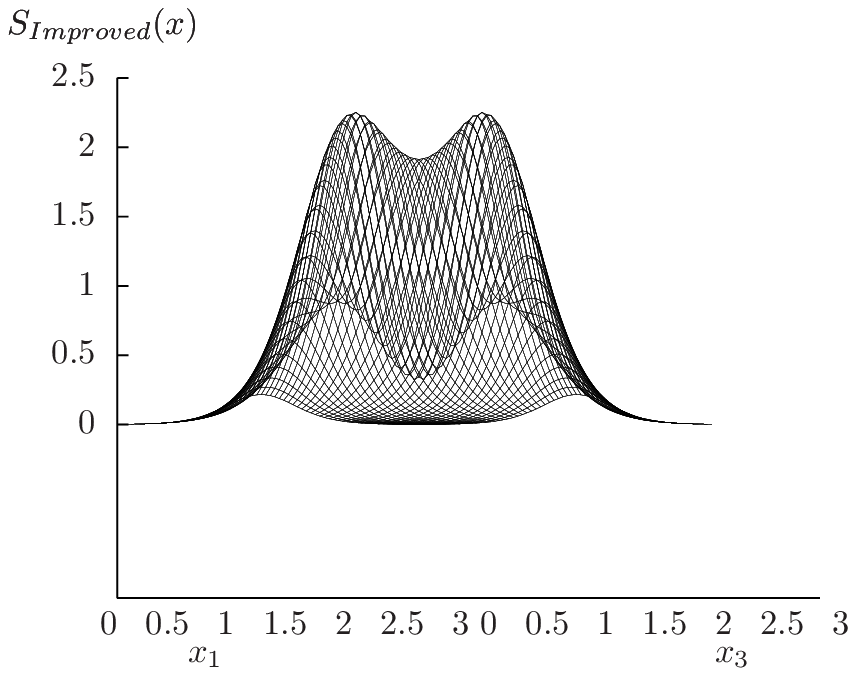}
h
\end{minipage}
\begin{minipage}{0.33\linewidth}
\includegraphics[angle=0,width=1.0\textwidth]{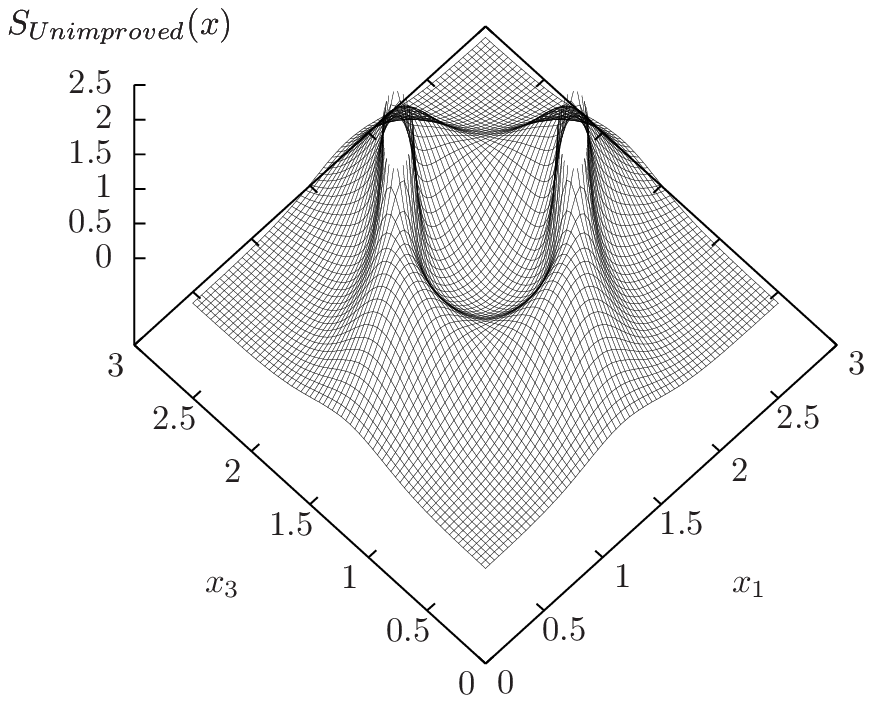}
c
\includegraphics[angle=0,width=1.0\textwidth]{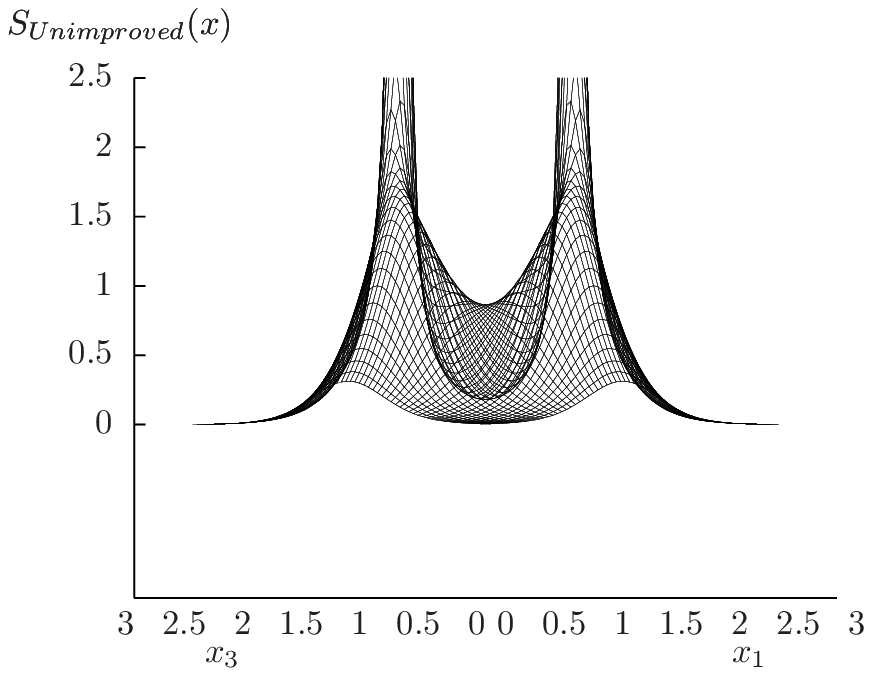}
f
\includegraphics[angle=0,width=1.0\textwidth]{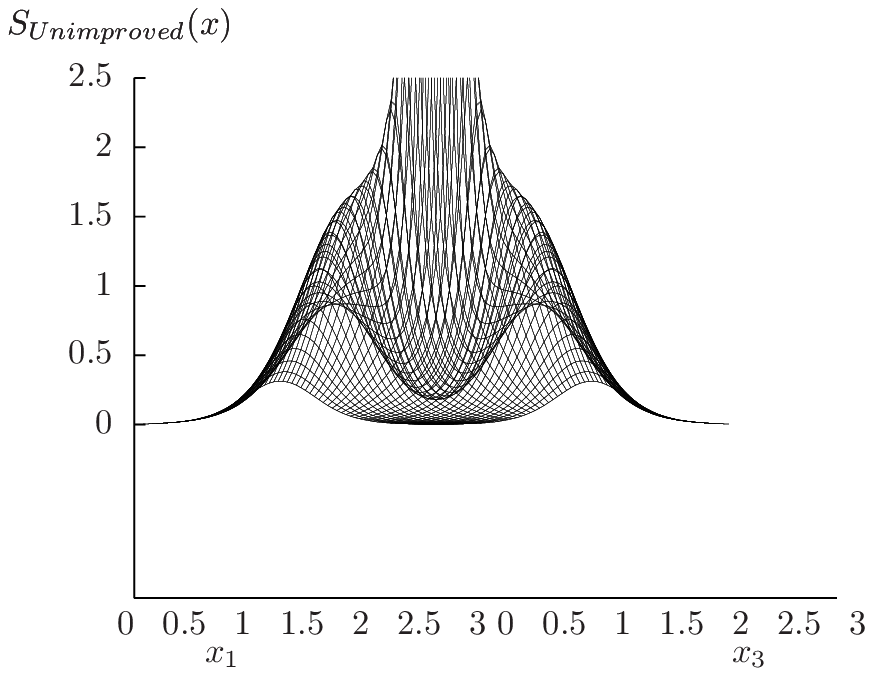}
i
\end{minipage}
}
\vs{2mm}
\caption{a,d,g: Action density of exact $Q=2$ solution. b,e,h: Action density of pseudo-ADHM
improved configuration. c,f,i: Action density of unimproved sum-ansatz configuration.
d,e,f: Horizontal view in direction ($dx_1=1, dx_3=1$). g,h,i: Horizontal view in direction 
($dx_1=-1, dx_3=1$).}
\label{fig:figExactQ2solution}
\end{figure}

\section{Summary and conclusions}
\label{sec:Conclusion}
In the present paper we have dealt with the question how approximate 
multi-(anti)caloron configurations can be constructed by using exact 
single (anti)caloron gauge fields as the underlying building blocks. 
We have discussed the first fundamental problem 
encountered when one attempts to carry over the well-known sum-ansatz, 
which can successfully be applied for the construction of not too dense 
multi-instanton systems, to the case of KvBLL calorons with non-trivial 
holonomy. The technical answer to this problem has already been used in 
our recent study~\cite{Gerhold:2006sk} of the caloron gas where we have 
found that it works relatively well even without further improvement.
This certainly applies only to the parameter range that we found to be relevant.
In the present paper we have proposed two improved superposition schemes to 
deal with difficulties arising when the caloron system is either too dense or 
when the calorons increasingly appear in a well-dissociated state.

The first fundamental problem, arising from the non-vanishing $A_4$ component 
due to the non-trivial holonomy of the considered calorons, can be overcome by 
superposing the caloron gauge fields in the algebraic gauge. This ansatz, 
however, implies the restriction that only calorons (and anticalorons) with 
identical holonomy become superposed, which has significant consequences when 
one is calculating observables like the Polyakov loop correlator in a 
semiclassical, caloron based simulation. This is due to the long-ranged 
correlations induced by this superposition scheme. Whether this 
restriction has a real physical background, meaning that multi-caloron 
configurations with variant single-caloron holonomies are actually 
suppressed, or whether it is simply a remnant of the imperfectness 
of the applied superposition scheme, remains unanswered here.

Secondly, we have dealt with the possible caloron-Dirac string interaction, 
which can easily create a tremendous action surplus when an additional caloron 
is inserted into the system in such a way that one of its monopoles comes too 
close to the Dirac string of another, well-dissociated caloron. We therefore 
have introduced the so-called DS-improved superposition scheme, which - loosely 
speaking - displaces the Dirac string before the vector potentials 
are superposed and moves the Dirac string back to the original place
afterwards. By comparison with the sum-ansatz we observed a greatly improved 
selfduality of the configurations obtained from this (Dirac-string improved) 
superposition scheme. The excess action is essentially reduced. This already 
cures some, increasingly unpleasant features one might encounter by employing 
the caloron gas model in a slightly modified physical regime with strongly 
separated monopoles.

We have checked that the DS-improved configurations are practically
stable under improved cooling ($\epsilon=0$ cooling) in contrast to the 
unimproved gauge fields. From these observations we conclude that even without 
cooling the configurations constructed within the DS-improved superposition 
scheme are very close to classical configurations.
This technique is also applicable in systems consisting of calorons {\it and} 
anticalorons which are not required to be close to any classical configuration 
and where cooling would be problematic as a tool to improve, in a second step, 
the configuration obtained by the sum-ansatz.
The only technical restriction for the DS-improvement scheme is that the constituents 
of the dissociated caloron containing the Dirac string have to be sufficiently 
far separated from each other.
We therefore conclude that the DS-improved superposition scheme might be the 
method of choice for the confinement phase if it turns out that the calorons 
should be more dissociated into monopoles (dyons) than we have found 
appropriate in our preceding paper~\cite{Gerhold:2006sk}.

We then presented another superposition scheme suitable to improve the 
classicality and selfduality or anti-selfduality of (anti)caloron 
superpositions in the case of overlapping calorons or monopole cores 
(of different calorons). Our so-called pseudo-ADHM ansatz is based on the 
ADHM formalism. Its main idea is to approximate the underlying ADHM data 
of the actual (anti-)selfdual multi-caloron configuration by combining 
single (anti)caloron ADHM data rather than attempting to improve the 
corresponding gauge field directly. Here, we simply unify the ADHM data 
of all involved calorons or anticalorons, which guarantees {\em (i)}~that 
the caloron center positions (subject to sampling) remain unchanged, 
{\em (ii)}~that the periodicity in time is exactly preserved, 
and {\em (iii)}~that an analytical expression in terms of the single 
(anti)caloron vector potentials including an easily calculable weighing 
factor exists. The price to pay is to abandon to solve the (nonlinear) 
ADHM constraint. For trivial holonomy and equal orientations in color 
space of all calorons being involved, this ansatz yields exactly selfdual 
configurations, boiling down to the well-known 't Hooft ansatz in the 
case of instantons. On the other hand, our pseudo-ADHM superposition scheme 
corresponds to Shuryak's ratio-ansatz when applied to instanton-antiinstanton 
pairs. Presenting a comparison of this improved superposition scheme 
with the sum-ansatz we have demonstrated that the selfduality of gauge 
fields created by the pseudo-ADHM  construction is significantly 
improved and the excess action essentially reduced.
This becomes a major advantage at small caloron or monopole distances. 

The overall structure of the improved configurations as represented 
by the action density is relatively stable under improved cooling. 
Overlapping monopoles are driven by cooling to slightly separate from each 
other, keeping their identity as monopoles. In contrast to these 
observations, the configurations obtained from the sum-ansatz are strongly 
changing their structure during the cooling process. 
This means that these gauge field configurations are rather far 
from classical ones. An interesting property of unimproved 
(sum-ansatz) configurations is that they are not protected against 
the merger of equal-charge monopoles into higher-charge
monopoles under cooling.  This is witnessed by the emergence of a 
typical volcano-like structure in the action density which is known 
to be a characteristic feature of exact charge $Q=2$ caloron 
solutions in the limit of approaching equal-charge monopoles. 
It should be kept in mind that the pseudo-ADHM ansatz is intentionally 
constructed in such a way to preserve the identity of the single monopoles.

As a final remark, the combination of the benefits of the DS-improvement 
and of the pseudo-ADHM technique would result in a much wider applicable 
version of the caloron gas model that we have started to develop in 
Ref.~\cite{Gerhold:2006sk}. However, since the weighing factor of the 
pseudo-ADHM ansatz changes the structure of the Dirac string, 
this has not been accomplished so far. This remains an important subject 
for further studies.

\section*{Acknowledgements}
We would like to thank Boris Martemyanov for introducing us to the 
practice of superposing KvBLL calorons with non-trivial holonomy in the algebraic gauge.
We are grateful to Falk Bruckmann for sharing his insights on KvBLL caloron
systems and for enlightening discussions and correspondence.
We appreciate the support from Pierre van Baal, who created the idea to
organize a meeting on caloron problems in Leiden in the summer 2005.
Both are thanked for critically reading the first version of this paper.
Furthermore, we thank D. Nogradi for making the data describing
the exact solution of Ref.~\cite{Bruckmann:2004nu} accessible to us.
We acknowledge financial support by the DFG through the Forschergruppe
FOR 465 / Mu 932/2-4.

\bibliographystyle{unsrtOWN}
\bibliography{ImprovedSupREVISED}

\end{document}